\documentclass[a4paper,11pt]{article}
\pdfoutput=1

\usepackage{jheppub}

\usepackage{amsmath,amssymb}
\usepackage{mathrsfs}

\usepackage{graphicx,tikz}
\usepackage[labelformat=simple]{subcaption}

\usepackage[format=plain]{caption}
\usepackage{tabularx}
\usepackage{enumitem}
\usepackage{cancel}
\usepackage{sectsty}
\allsectionsfont{\boldmath}

\providecommand{\d}{\mathop{}\!\mathrm{d}}
\renewcommand{\d}{\mathop{}\!\mathrm{d}}

\newcommand*{\Ord}{\mathcal{O}}
\newcommand*{\RR}{\mathbb{R}}

\newcommand*{\ie}{i.e.\ }

\title{Supersymmetric black hole non-uniqueness in five dimensions}

\author{Veronika Breunh\"older}
\author{and James Lucietti}
\affiliation{School of Mathematics and Maxwell Institute for
  Mathematical Sciences,\\University of Edinburgh, The King's
  Buildings, Edinburgh, EH9 3FD, UK}
\emailAdd{v.breunhoelder@ed.ac.uk}
\emailAdd{j.lucietti@ed.ac.uk}

\abstract{We present a systematic study of the moduli space of
  asymptotically flat, supersymmetric and biaxisymmetric black hole
  solutions to five-dimensional minimal supergravity. Previously, it
  has been shown that such solutions must be multi-centred solutions
  with a Gibbons--Hawking base. In this paper we perform a full
  analysis of three-centred solutions with a single black hole, for
  which there are seven regular black hole solutions.  We find that
  four of these can have the same conserved charges as the BMPV black
  hole. These consist of a black lens with $L(3,1)$ horizon topology
  and three distinct families of spherical black holes with nontrivial
  topology outside the horizon. The former provides the first example
  of a nonspherical black hole with the same conserved charges as the
  BMPV black hole.  Moreover, of these four solutions, three can have
  a greater entropy than the BMPV black hole near the BMPV upper spin
  bound. One of these is a previously known spherical black hole with
  nontrivial topology and the other two are new examples of a
  spherical black hole with nontrivial topology and an $L(3,1)$ black
  lens.}

\begin{document} 
\maketitle
\flushbottom

\section{Introduction}

It is well known that higher dimensional black holes are not uniquely specified by their asymptotic charges. The first example which demonstrated this was an asymptotically flat black hole solution to the vacuum Einstein equations with $S^1\times S^2$ horizon topology,  known as the black ring~\cite{Emparan:2001wn}.  Together with the higher dimensional analogue of the Kerr solution found by Myers and Perry~\cite{Myers:1986un}, this was sufficient to establish that even asymptotically flat vacuum black holes are not uniquely fixed by their asymptotic charges. It is by now expected that the moduli space of higher dimensional black holes is vast and complicated, although exact black hole solutions to the Einstein equations remain elusive~\cite{Emparan:2008eg, Hollands:2012xy}. 

What is perhaps less well known is that even asymptotically flat, supersymmetric black holes exhibit the phenomenon of black hole non-uniqueness. The first example which demonstrated this (with a connected horizon) was a supersymmetric black ring in five-dimensional $U(1)^3$-supergravity, which carries local dipole charges that are not fixed by the conserved charges~\cite{Elvang:2004ds}.  However, this solution can never have the same conserved charges as the spherical topology BMPV black hole~\cite{Breckenridge:1996is}, since the former has unequal nonzero angular momenta whereas the latter has equal angular momenta (with respect to orthogonal 2-planes at infinity). Therefore, the existence of this black ring solution was not in conflict with the string theory computation of the entropy of the BMPV black hole, which counts quantum states with the same conserved charges as the black hole~\cite{Strominger:1996sh, Breckenridge:1996is}.

More recently, another more basic mechanism leading to black hole non-uniqueness has been discovered:  {\it nontrivial spacetime topology.} In particular, asymptotically flat, supersymmetric black holes with spherical horizon topology and a nontrivial cycle  of $S^2$-topology in the exterior spacetime can be constructed in five-dimensional minimal supergravity~\cite{Kunduri:2014iga}. Furthermore, these  solutions can carry the same asymptotic charges as the BMPV black hole, thus demonstrating supersymmetric black hole non-uniqueness even for spherical topology black holes~\cite{Kunduri:2014iga}. Strikingly, it was found that their entropy can be larger than that of the BMPV black hole near the BMPV upper spin bound~\cite{Horowitz:2017fyg}.\footnote{This has also been observed for multi-black rings and black holes~\cite{Gauntlett:2004qy, Crichigno:2016lac} and asymptotically AdS black holes~\cite{Bena:2011zw}.} This is particularly interesting as it is poses a challenge to the aforementioned microscopic entropy computation of the BMPV black hole. It remains to be understood how this apparent conflict may be resolved.

The black holes with nontrivial spacetime topology belong to the same
class of solutions as the bubbling ``microstate'' geometries~\cite{Bena:2005va}. These are asymptotically flat smooth soliton geometries with noncontractible 2-cycles, or ``bubbles'', and are constructed from multi-centred solutions with a Gibbons--Hawking base. Indeed, it was shown that the black holes with nontrivial topology~\cite{Kunduri:2014iga} may be interpreted as black holes sitting inside a bubbling soliton geometry~\cite{Horowitz:2017fyg}. In fact, another novel type of black hole has been discovered in this class of solutions: a black hole with lens space topology known as a black lens~\cite{Kunduri:2014kja, Kunduri:2016xbo}. This black lens has $L(2,1) \cong 
\mathbb{RP}^3$ horizon topology and always possesses unequal angular momenta, so may never carry the same conserved charges as the BMPV black hole. It appears this is also the case for the more general class of black lenses with $L(p,1)$ topology subsequently constructed~\cite{Tomizawa:2016kjh}.

Recently, a complete classification of asymptotically flat,
supersymmetric and biaxisymmetric black hole (and soliton) solutions
to five-dimensional minimal supergravity has been
derived~\cite{Breunholder:2017ubu}. This reveals a vast moduli space
of spherical black holes, black rings and black lenses in spacetimes with noncontractible 2-cycles. This includes all the above examples, but also reveals infinitely many more black hole solutions for each horizon topology. Given this, one expects the above example of supersymmetric black hole non-uniqueness to be merely the tip of the iceberg.  Indeed, several natural questions present themselves: Are there other black hole solutions with the same charges as the BMPV black hole? Do any of these have nonspherical horizon topology? Do any of these have greater entropy than the BMPV black hole?

The purpose of this paper is to use the recent
classification~\cite{Breunholder:2017ubu} to perform a systematic
study of the moduli space of supersymmetric black holes of
five-dimensional minimal supergravity. In particular, we wish to study
in detail the moduli space of solutions with the same conserved
charges as the BMPV black hole.  Even given the explicit
classification this is technically challenging and at present out of
reach. This is essentially due to the fact the general solution
depends on many parameters that are subject to polynomial equations
and several inequalities, arising from smoothness and causality, which
are difficult to disentangle.  The classification shows that the
solutions must have a Gibbons--Hawking base with harmonic functions of
multi-centred type and that the number of centres $n$ is the number of
black holes plus the number of corners of the orbit space $\hat{M}=
M/(\mathbb{R}\times U(1)^2$) (this is the number of fixed points of
the $U(1)^2$-action on the spacetime). Thus even restricting to single black hole solutions, as we will, leaves an infinite class of solutions enumerated by the number of centres $n$. We will investigate this question for the lowest nontrivial values of $n$. For $n=1$ there is only the BMPV black hole, whereas for $n=2$ there is only the black ring~\cite{Elvang:2004rt} and $L(2,1)$ black lens~\cite{Kunduri:2014kja}, which as noted above never have the same charges as BMPV.   We will content ourselves with a study of single black hole solutions with $n=3$ centres and find that even this is rich enough to answer the above questions. 

Our main results are as follows.
We find that there are four $3$-centred single black hole solutions
that can have the same conserved charges as the BMPV black hole, one
of which is the spherical black hole with nontrivial topology
mentioned above~\cite{Kunduri:2014iga, Horowitz:2017fyg}.
Furthermore, three of these may have an entropy greater than that of
the BMPV solution near the BMPV upper spin bound, one of which is the
previously known case~\cite{Horowitz:2017fyg}. The two new cases
correspond to a distinct family of spherical black holes with nontrivial topology (with no bubble in the exterior, only disc topology surfaces ending on the horizon), and an $L(3,1)$ black lens that is a distinct solution to the previously studied $L(p,1)$ black lens~\cite{Tomizawa:2016kjh}.  In particular, our $L(3,1)$ black lens provides the first example of a black hole with nonspherical topology and the same conserved charges as the BMPV black hole.  

This paper is organised as follows. In section
\ref{sec:form_of_solutions} we briefly review the classification of asymptotically flat, supersymmetric and biaxisymmetric black holes of five-dimensional minimal supergravity, focusing on the $n=3$ centred solutions. In section \ref{sec:equal_J} we analyse the moduli space of solutions which can have the same conserved charges as the
BMPV black hole. In section \ref{sec:comparison} we present a comparison of the entropies of these black holes. In section \ref{sec:discussion} we  discuss our results. We also give an Appendix with details on smoothness and causality of our solutions.

\section{Supersymmetric solutions of five-dimensional minimal
  supergravity}\label{sec:form_of_solutions}

Asymptotically flat, supersymmetric and biaxisymmetric (\ie possessing a
$U(1)^2$-symmetry that commutes with supersymmetry) black hole and soliton solutions to five-dimensional minimal supergravity
have recently been classified \cite{Breunholder:2017ubu}. The proof is based on the
well known local classification of supersymmetric solutions \cite{Gauntlett:2002nw},
as well as the classification of the possible near-horizon geometries
\cite{Reall:2002bh}.  We now give a brief
overview of these solutions. 

The bosonic fields consist of a
metric and Maxwell field, 
\begin{align}
  \label{g}
  \d s^2 &= -f^2( \d t + \omega)^2 + f^{-1} h\,,\\
  \label{F}
  F & = \d A = \frac{\sqrt{3}}{2} \d \left[ f( \d t + \omega) - K
      H^{-1} (\d \psi+ \chi)  - \xi \right]  \,,
\end{align} 
where $h$ is a Gibbons--Hawking metric,
\begin{equation}\label{GH}
h = H^{-1} (\d\psi + \chi)^2 + H (\d r^2 + r^2 \d\Omega_2^2) \, .
\end{equation}
The solution is completely determined by four harmonic
functions $H$, $K$, $L$, and $M$ on the $\RR^3$-base of $h$ (here
written in spherical coordinates) via
\begin{gather}
  f^{-1} = H^{-1} K^2 + L \, , \label{Ldef} \\
  \omega = \omega_\psi(\d\psi + \chi) + \hat{\omega} \,, \qquad
  \omega_\psi = H^{-2} K^3 + \frac{3}{2}H^{-1}KL + M \,
                , \label{Mdef}
\end{gather}
and with the remaining 1-forms $\chi$, $\hat\omega$ and $\xi$ (all on $\RR^3$) given by
\begin{equation}\label{1-forms}
  \star_3 \d\chi = \d H\,, \qquad
  \star_3 \d\hat\omega = H \d M - M \d H + \frac{3}{2}(K \d L -
                           L \d K) \,,\qquad
  \star_3 \d\xi = - \d K\,.
\end{equation} 
One  of the main results of the classification~\cite{Breunholder:2017ubu} is that regularity of the spacetime both on and outside the event horizon implies that the harmonic functions must be of multi-centred type,
\begin{equation}\label{harmonicfunc}
    H = \sum_{i=0}^{n-1} \frac{h_i}{r_i}\,,\qquad
    K = \sum_{i=0}^{n-1} \frac{k_i}{r_i}\,,\qquad
    L = 1 + \sum_{i=0}^{n-1} \frac{\ell_i}{r_i}\,,\qquad
    M = m + \sum_{i=0}^{n-1} \frac{m_i}{r_i}\,,
\end{equation}
where $r_i=\sqrt{r^2+a_i^2-2ra_i\cos\theta}$ and $n\geq 1$ is the
number of centres. The centres are positioned on the $z$-axis of the
$\RR^3$ at $z=a_i$ where $a_i \neq a_j$ for $i\neq j$. Furthermore,
the number of centres $n=k+l$ where $k$ is the number of connected
components of the event horizon and $l$ the number of corners of the
orbit space (points where both rotational Killing fields vanish).  The parameters
$\{h_i,k_i,\ell_i,m,m_i,a_i\}_{0\leq i < n}$ must satisfy a number of
constraint equations and inequalities arising from regularity of the
spacetime, summarised in~\cite{Breunholder:2017ubu}. In general, these
constraints are a complicated set of polynomial equations and
inequalities, which makes studying solutions in more depth a difficult
task (although see \cite{Avila:2017pwi} for recent progress on the
pure soliton case).

For single black hole solutions ($k=1$) with
$n\leq 3$ the constraints are exactly solvable. The $n=1$ solution reduces to the BMPV black
hole; for $n=2$ there are two possible
solutions corresponding to the $L(2,1)$ black lens
\cite{Kunduri:2014kja} and the supersymmetric black ring
\cite{Elvang:2004rt}. Neither of these can
have the same conserved charges as the BMPV black hole. In this paper we will consider solutions with $n=3$; in this case it is already known that there is at least one solution which may have the same charges as the BMPV black hole~\cite{Kunduri:2014iga, Horowitz:2017fyg}.

For a single black hole solution with $n=3$ the harmonic functions
\eqref{harmonicfunc} are~\cite{Breunholder:2017ubu}\footnote{We have exploited a gauge freedom in
  the harmonic functions to set $k_0=0$.}
\begin{equation}
\begin{aligned}
    H &= \frac{h_0}{r} + \frac{h_1}{r_1} + \frac{h_2}{r_2}\,,\qquad
    &K &= \frac{k_1}{r_1} + \frac{k_2}{r_2}\,,\\
    L &= 1 + \frac{\ell_0}{r} - \frac{h_1 k_1^2}{r_1} - \frac{h_2k_2^2}{r_2}\,,\qquad
    &M &= -\frac{3}{2}(k_1+k_2) + \frac{m_0}{r} + \frac{k_1^3}{2r_1} + \frac{k_2^3}{2r_2}\,,
\end{aligned}
\end{equation}
where we have fixed $r=0$ to be the position of the horizon and $r_{1}=0$ and $r_{2}=0$ are fixed points of the biaxial symmetry. Furthermore, $h_0+h_1+h_2 =1$, $h_{1,2} = \pm 1$, $h_0 \in \mathbb{Z}$, and
the remaining parameters are subject to the constraints
\begin{equation}
  \frac{3}{2}(k_i-h_i(k_1+k_2))+\frac{-\tfrac{1}{2}h_0k_i^3
  + \tfrac{3}{2}k_i\ell_0 + h_im_0}{|a_i|}+(-1)^i
  \frac{(h_2k_1-h_1k_2)^3}{2|a_2-a_1|} = 0\,,\qquad i=1,2\label{conds}
\end{equation}
and inequalities
\begin{align}
  -h_0^2m_0^2 + h_0\ell_0^3 &> 0\,, \label{inequ_horizon}\\
  h_i\left(1+\frac{-h_0k_i^2+\ell_0}{|a_i|}\right) -
  \frac{h_1h_2(h_2k_1-h_1k_2)^2}{|a_2-a_1|} &> 0\,,\qquad i=1,2\,.
                                              \label{inequs_centres}
\end{align}
 which arise from the appropriate smoothness conditions at the three centres.
Cross-sections of the horizon have area
\begin{equation}\label{AH}
  A_H = 16\pi^2 \sqrt{
    \ell_0^3 h_0 -  m_0^2h_0^2} \,,
\end{equation}
and the asymptotic charges of the solution are
\begin{align}
  Q &=  \frac{2}{\sqrt{3}} M_{\text{\tiny ADM}} = 2\sqrt{3}\pi\Big(-h_1 k_1^2 - h_2 k_2^2
      + (k_1+k_2)^2+\ell_0\Big)\,,\label{Q}\\
  J_\psi &= 2\pi \Big(\tfrac{1}{2}(k_1^3+k_2^3) +
           (k_1+k_2)^3-\tfrac{3}{2}(k_1+k_2)(h_1k_1^2+h_2k_2^2-\ell_0)
           + m_0 \Big)\,,\label{Jpsi}\\
  J_\phi &= 3\pi \Big( a_1(k_1-h_1(k_1+k_2))+a_2(k_2-h_2(k_1+k_2)) \Big)\,.\label{Jphi}
\end{align}
Finally, for the solution to be smooth and stably causal, we require
\begin{equation}\label{smoothness_causality}
  K^2+HL>0\,, \qquad g^{tt}<0  \; ,
\end{equation}
in the domain of outer communication $r>0$.

It is worth noting that, in general,
(\ref{conds}--\ref{inequs_centres}) are not sufficient to ensure
positivity of the mass of the solution. In fact, numerical checks suggest
that $M_{\text{\tiny ADM}}>0$ is implied by (\ref{conds}--\ref{inequs_centres}) together with \eqref{smoothness_causality} (as must be the case from the BPS bound). Conversely, our checks also support the following conjecture: (\ref{conds}--\ref{inequs_centres})  together with $M_{\text{\tiny ADM}}>0$ guarantee \eqref{smoothness_causality} are obeyed.  If true, this would greatly simplify the study of the moduli space of supersymmetric black holes. In the Appendix, we present some numerical checks which support this conjecture.
 In any case, we will explicitly add
\begin{equation}\label{posmass}
  M_{\text{\tiny ADM}} = 3\pi\Big(-h_1 k_1^2 - h_2 k_2^2
      + (k_1+k_2)^2+\ell_0\Big) > 0
\end{equation}
to our set of constraints on the parameters and verify
\eqref{smoothness_causality} numerically. In fact, if $h_0=3,
h_1=h_2=-1$ the positive mass inequality \eqref{posmass} and
smoothness condition $K^2+HL>0$ are automatically satisfied as a consequence of (\ref{conds}--\ref{inequs_centres}) (see Appendix for proof).

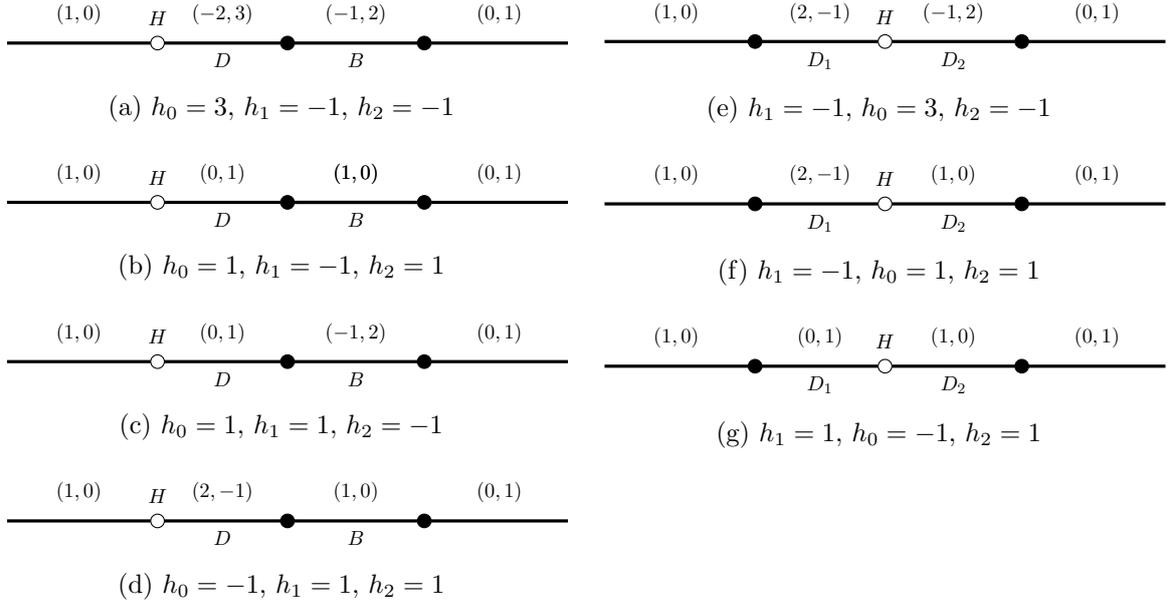
\begin{figure}[t!]
\centering
\begin{minipage}[t]{.48\textwidth}
\centering
\begin{subfigure}{\linewidth}
\begin{tikzpicture}[scale=0.9, every node/.style={scale=0.7}]
  \draw[very thick](-4,0)--(-1.9,0)node[midway,above=.2cm]{$(1,0)$};
  \draw[very thick](-1.7,0)--(0.0,0)node[midway,above=0.2cm]{$(-2,3)$}%
  node[midway,below=.05cm]{$D$}; 
  \draw[very thick](0.2,0)--(2.0,0)node[midway, above= 0.2cm]{$(-1,2)$}%
  node[midway,below=.05cm]{$B$};
  \draw[very thick](2.2,0)--(4.2,0)node[midway, above= 0.2cm]{$(0,1)$};
  \draw[fill=white] (-1.8,0) circle [radius=.1] node[above=.2cm]{$H$};
  \draw[fill=black] (0.1,0) circle [radius=.1]; 
  \draw[fill=black] (2.1,0) circle [radius=.1];
\end{tikzpicture}
\caption{$h_0=3$, $h_1=-1$, $h_2=-1$}
\label{h=3-1-1}
\end{subfigure}
\\\vspace{1em}
\begin{subfigure}{\linewidth}
\begin{tikzpicture}[scale=0.9, every node/.style={scale=0.7}]
  \draw[very thick](-4,0)--(-1.9,0)node[midway,above=.2cm]{$(1,0)$};
  \draw[very thick](-1.7,0)--(0.0,0)node[midway,above=.2cm]{$(0,1)$}%
  node[midway,below=.05cm]{$D$};
  \draw[very thick](0.2,0)--(2.0,0)node[midway,above=.2cm]{$(1,0)$}%
  node[midway,below=.05cm]{$B$};
  \draw[very thick](0.2,0)--(2.0,0)node[midway,above=.2cm]{$(1,0)$};
  \draw[very thick](2.2,0)--(4.2,0)node[midway,above=.2cm]{$(0,1)$};
  \draw[fill=white] (-1.8,0) circle [radius=.1] node[above=.2cm]{$H$};
  \draw[fill=black] (0.1,0) circle [radius=.1];
  \draw[fill=black] (2.1,0) circle [radius=.1];
\end{tikzpicture}
\caption{$h_0=1$, $h_1=-1$, $h_2=1$}
\label{h=1-11_1}
\end{subfigure}
\\\vspace{1em}
\begin{subfigure}{\linewidth}
\begin{tikzpicture}[scale=0.9, every node/.style={scale=0.7}]
  \draw[very thick](-4,0)--(-1.9,0)node[midway,above=.2cm]{$(1,0)$};
  \draw[very thick](-1.7,0)--(0.0,0)node[midway,above=.2cm]{$(0,1)$}%
  node[midway,below=.05cm]{$D$};
  \draw[very thick](0.2,0)--(2.0,0)node[midway,above=.2cm]{$(-1,2)$}%
  node[midway,below=.05cm]{$B$};
  \draw[very thick](2.2,0)--(4.2,0)node[midway,above=.2cm]{$(0,1)$};
  \draw[fill=white] (-1.8,0) circle [radius=.1] node[above=.2cm]{$H$};
  \draw[fill=black] (0.1,0) circle [radius=.1];
  \draw[fill=black] (2.1,0) circle [radius=.1];
\end{tikzpicture}
\caption{$h_0=1$, $h_1=1$, $h_2=-1$}
\label{h=11-1}
\end{subfigure}
\\\vspace{1em}
\begin{subfigure}{\linewidth}
\begin{tikzpicture}[scale=0.9, every node/.style={scale=0.7}]
  \draw[very thick](-4,0)--(-1.9,0)node[midway,above=.2cm]{$(1,0)$};
  \draw[very thick](-1.7,0)--(0.0,0)node[midway,above=.2cm]{$(2,-1)$}%
  node[midway,below=.05cm]{$D$};
  \draw[very thick](0.2,0)--(2.0,0)node[midway,above=.2cm]{$(1,0)$}%
  node[midway,below=.05cm]{$B$};
  \draw[very thick](2.2,0)--(4.2,0)node[midway,above=.2cm]{$(0,1)$};
  \draw[fill=white] (-1.8,0) circle [radius=.1] node[above=.2cm]{$H$};
  \draw[fill=black] (0.1,0) circle [radius=.1];
  \draw[fill=black] (2.1,0) circle [radius=.1];
\end{tikzpicture}
\caption{$h_0=-1$, $h_1=1$, $h_2=1$}
\label{h=-111_1}
\end{subfigure}
\end{minipage}\hspace{.04\textwidth}%
\begin{minipage}[t]{.48\textwidth}
\centering
\begin{subfigure}{\linewidth}
  \begin{tikzpicture}[scale=0.9, every node/.style={scale=0.7}]
    \draw[very thick](-4,0)--(-1.9,0)node[midway,above=.2cm]{$(1,0)$};
    \draw[very
    thick](-1.7,0)--(0.0,0)node[midway,above=.2cm]{$(2,-1)$}%
    node[midway,below=.05cm]{$D_1$};
    \draw[very
    thick](0.2,0)--(2.0,0)node[midway,above=.2cm]{$(-1,2)$}%
    node[midway,below=.05cm]{$D_2$};
    \draw[very thick](2.2,0)--(4.2,0)node[midway,above=.2cm]{$(0,1)$};
    \draw[fill=black] (-1.8,0) circle [radius=.1] ;
    \draw[fill=white] (0.1,0) circle [radius=.1] node[above=.2cm]{$H$};
    \draw[fill=black] (2.1,0) circle [radius=.1];
\end{tikzpicture}
\caption{$h_1=-1$, $h_0=3$, $h_2=-1$}
\label{h=-13-1}
\end{subfigure}
\\\vspace{1em}
\begin{subfigure}{\linewidth}
\begin{tikzpicture}[scale=0.9, every node/.style={scale=0.7}]
  \draw[very thick](-4,0)--(-1.9,0)node[midway,above=.2cm]{$(1,0)$};
  \draw[very thick](-1.7,0)--(0.0,0)%
  node[midway,above=.2cm]{$(2,-1)$}node[midway,below=.05cm]{$D_1$};
  \draw[very thick](0.2,0)--(2.0,0)%
  node[midway,above=.2cm]{$(1,0)$}node[midway,below=.05cm]{$D_2$};
  \draw[very thick](2.2,0)--(4.2,0)node[midway,above=.2cm]{$(0,1)$};
  \draw[fill=black] (-1.8,0) circle [radius=.1];
  \draw[fill=white] (0.1,0) circle [radius=.1] node[above=.2cm]{$H$};
  \draw[fill=black] (2.1,0) circle [radius=.1];
\end{tikzpicture}
\caption{$h_1=-1$, $h_0=1$, $h_2=1$}
\label{h=-111_2}
\end{subfigure}
\\\vspace{1em}
\begin{subfigure}{\linewidth}
\begin{tikzpicture}[scale=0.9, every node/.style={scale=0.7}]
  \draw[very thick](-4,0)--(-1.9,0)node[midway,above=.2cm]{$(1,0)$};
  \draw[very thick](-1.7,0)--(0.0,0)%
  node[midway,above=.2cm]{$(0,1)$}node[midway,below=.05cm]{$D_1$};
  \draw[very thick](0.2,0)--(2.0,0)%
  node[midway,above=.2cm]{$(1,0)$}node[midway,below=.05cm]{$D_2$};
  \draw[very thick](2.2,0)--(4.2,0)node[midway,above=.2cm]{$(0,1)$};
  \draw[fill=black] (-1.8,0) circle [radius=.1]; \draw[fill=white]
  (0.1,0) circle [radius=.1] node[above=.2cm]{$H$}; \draw[fill=black]
  (2.1,0) circle [radius=.1] ;
\end{tikzpicture}
\caption{$h_1=1$, $h_0=-1$, $h_2=1$}
\label{h=1-11_2}
\end{subfigure}
\end{minipage}
\caption{Rod diagrams for all seven 3-centred single black hole
  solutions. \protect\subref{h=3-1-1}--\protect\subref{h=-111_1} have the horizon at
  the first centre ($0<a_1<a_2$); \protect\subref{h=-13-1}--\protect\subref{h=1-11_2}
  have a central horizon ($a_1<0<a_2$). The numbers above each axis
  rod specify the biaxial Killing vector that vanishes on the rod with respect to
  the basis $(v_-, v_+)$ where $v_\pm = \partial_\phi \mp \partial_\psi$.}
\label{rod_structures}
\end{figure}

As mentioned above, we have fixed the position of the horizon to be
the origin of the $\RR^3$-base, $r=0$. This can be done without loss
of generality, however, we need to distinguish between the different
possible orders of the centres along the $z$-axis: $0<a_1<a_2$, which
corresponds to a horizon at the first centre, or $a_1<0<a_2$, in which
case the horizon is positioned between the other two
centres.\footnote{A potential horizon at the third centre is
  equivalent to one at the first centre, as they are related by a reflection symmetry along the $z$-axis.} With the above
constraints on $(h_0,h_1,h_2)$, this results in seven distinct rod
structures, depicted in Figure \ref{rod_structures}.  These lead to seven distinct regular black hole solutions.

The rod structure determines the spacetime and horizon topology. If $0<a_1<a_2$ (Figure
\ref{rod_structures}\subref{h=3-1-1}--\subref{h=-111_1}), the axis
rods $[0,a_1], [a_1, a_2]$ correspond to a disc topology surface $D$ ending on the horizon and an $S^2$-topology surface 
$B$ (bubble), respectively. If $a_1<0<a_2$ (Figure
\ref{rod_structures}\subref{h=-13-1}--\subref{h=1-11_2}) the axis rods both
correspond to discs $D_1$ and $D_2$ ending on the horizon
at $r=0$. These noncontractible 2-cycles correspond to the fixed points of
the rotational Killing vector fields along the axis. For each axis rod
$I$ corresponding to a 2-cycle $C$ (bubble or disc) one can define a magnetic
potential via $\d\Phi_I = \iota_{v_I} F$, where $v_I$ is the Killing
vector vanishing on $I$ and the integration constants are fixed such that
$\Phi_I$ vanishes asymptotically~\cite{Kunduri:2013vka}. By definition, the magnetic
potentials take constant values $q_{C} \equiv \Phi_I|_I$ on the
corresponding axis rods. For the 3-centred solutions in Figure
\ref{rod_structures}, these ``dipole charges'' $q_C$ are
\begin{align}
  q_D &= \sqrt{3}h_0(k_1+k_2)\,, & q_B &=
  \sqrt{3}(k_2-h_2(k_1+k_2))\, &\text{for } 0<a_1<a_2\,,\label{potentials1}\\
  q_{D_1} &= \sqrt{3}(-k_1 +h_1(k_1+k_2))\,,& q_{D_2} &=
  \sqrt{3}(k_2-h_2(k_1+k_2))\, &\text{for } a_1<0<a_2\,.\label{potentials2}
\end{align}
The nontrivial 2-cycle structure is supported by
the corresponding magnetic fluxes. Note that the dipole charges $q_C$ are not conserved
charges (they are not related to a symmetry of the
solution).

Let us finally consider the horizon topologies of the solutions: if $h_0=\pm 1$ the
horizon is of $S^3$ topology, whereas in general $h_0=p$ for $p\in \mathbb{Z}$
corresponds to a horizon of topology $L(p,1) \cong S^3/ \mathbb{Z}_p$.\footnote{The case $h_0=0$ corresponds to a horizon of topology
  $S^1\times S^2$, however this cannot occur in this class of solutions.}
Thus, the solutions defined by Figures \ref{h=3-1-1} and \ref{h=-13-1}
are black lenses with horizon topology $L(3,1)$, whereas Figures
\ref{h=1-11_1}--\subref{h=-111_1} and Figures
\ref{h=-111_2}--\subref{h=1-11_2} are spherical ($S^3$) black
holes with nontrivial topology in the exterior. Note that only two of these solutions have been previously studied: Figure \ref{h=1-11_1} is the black hole with bubble solution in
\cite{Kunduri:2014iga,Horowitz:2017fyg}, and Figure \ref{h=3-1-1}  is the black lens
solution in \cite{Tomizawa:2016kjh}.

\begin{figure}[b!]
\centering
%%%%% 1st picture:
\begin{subfigure}{\linewidth}
\centering
\begin{tikzpicture}[scale=1, every node/.style={scale=0.7}]
  \draw[very thick](-4,0)--(-1.9,0)node[midway,above=.2cm]{$(1,0)$};
  \draw[very thick](-1.7,0)--(0.0,0)node[midway,above=0.2cm]{$(0,1)$};
  \draw[very thick](0.2,0)--(2.0,0)node[midway,above= 0.2cm]{$(-1,2)$};
  \draw[very thick](2.2,0)--(4.2,0)node[midway,above= 0.2cm]{$(0,1)$};
  \draw[fill=black] (-1.8,0) circle [radius=.1];
  \draw[fill=black] (0.1,0) circle [radius=.1]; 
  \draw[fill=black] (2.1,0) circle [radius=.1];
\end{tikzpicture}
\caption{$h_0=1, h_1=1, h_2=-1$}
\label{sol:h11-1}
\end{subfigure}
\vspace{2em}
%%%%% 2nd picture:
\begin{subfigure}{\linewidth}
\centering
\begin{tikzpicture}[scale=1, every node/.style={scale=0.7}]
  \draw[very thick](-4,0)--(-1.9,0)node[midway,above=.2cm]{$(1,0)$};
  \draw[very thick](-1.7,0)--(0.0,0)node[midway,above=.2cm]{$(0,1)$};
  \draw[very thick](0.2,0)--(2.0,0)node[midway,above=.2cm]{$(1,0)$};
  \draw[very thick](2.2,0)--(4.2,0)node[midway,above=.2cm]{$(0,1)$};
  \draw[fill=black] (-1.8,0) circle [radius=.1];
  \draw[fill=black] (0.1,0) circle [radius=.1];
  \draw[fill=black] (2.1,0) circle [radius=.1];
\end{tikzpicture}
\caption{$h_0=1, h_1=-1, h_2=1$}
\label{sol:h1-11}
\end{subfigure}
\caption{Inequivalent rod structures for 3-centred solitons.}
\label{fig:solitons}
\end{figure}
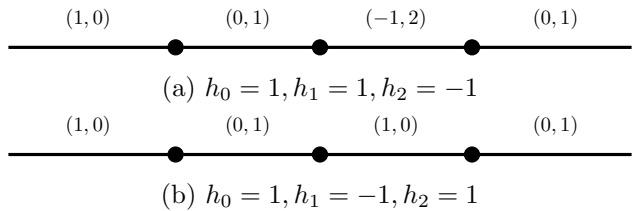

We will also consider the soliton spacetimes in this class, i.e.,
asymptotically flat everywhere smooth spacetimes with no black hole.
These are constructed as above except the boundary condition at $r=0$
is chosen to correspond to a corner of the orbit space rather than a horizon. The resulting constraints on the parameters are as above except \eqref{inequ_horizon} is now replaced by
\begin{equation}
\ell_0=0, \qquad m_0=0, \qquad h_0 \left( 1- \frac{h_1 k_1^2}{a_1} - \frac{h_2 k_2^2}{a_2} \right)>0 \;,   \label{solitonreg}
\end{equation}
where now without loss of generality we may take $0<a_1<a_2$~\cite{Breunholder:2017ubu}. The rod structures are more constrained in this case and there are only two inequivalent soliton solutions as depicted in Figure \ref{fig:solitons}. 
It should be noted that for the solitons the 2-cycles $C_1$ and $C_2$ corresponding to the two axis rods $[0,a_1]$ and $[a_1, a_2]$ respectively are both topologically $S^2$.

\section{Comparison with the BMPV black hole: Equal angular
  momenta}\label{sec:equal_J}

We now determine whether the conserved charges of the solutions described in the previous section can coincide with those of the
BMPV black hole. The BMPV solution is a rotating black hole with equal
angular momenta
\begin{equation}\label{equalJpm}
|J_+|=|J_-|   \; ,
\end{equation}
defined with respect to two orthogonal $U(1)^2$-Killing fields at
infinity (normalised to $2\pi$ period), say $v_\pm = \partial_\phi \mp \partial_\psi$.  Equality of $J_+$ and $J_-$ is only defined up to
a sign, as there is no fixed relative orientation of the corresponding orthogonal 2-planes. $J_+$ and $J_-$ can be related to the angular momenta defined with
respect to the Euler angles $(\psi,\phi)$ of the $S^3$ at infinity via
$J_\pm=J_\phi \mp J_\psi$. Then
\eqref{equalJpm} translates to
\begin{equation}\label{equalJ}
  J_\phi = 0 \qquad \text{or}\qquad J_\psi =0\,.
\end{equation}
An analysis of the constraints \eqref{conds}--\eqref{inequs_centres} and \eqref{posmass} on the parameter space for the 3-centred solutions reveals that some of
them allow for \eqref{equalJ}, whereas others always have
strictly distinct charges to BMPV. The results of this analysis are
summarised in Figure \ref{table}, which gives a list of all 3-centred
solutions, and whether or not \eqref{equalJ} is allowed. We will discuss the four solutions for which \eqref{equalJ} is possible in more detail below. For simplicity we will refer to the three
solutions with an $S^3$-horizon that allow \eqref{equalJ} as ``spherical black hole (with nontrivial topology) I, II, III'', and to the $L(3,1)$ black hole that allows for  \eqref{equalJ} simply
as ``black lens''. We emphasise though that the latter is a distinct
solution to the previously studied $L(3,1)$ black hole depicted in
Figure \ref{h=3-1-1}, for which \eqref{equalJ} is never possible~\cite{Tomizawa:2016kjh}.  

For each of the above solutions with equal angular momenta we have verified numerically that the smoothness and causality conditions \eqref{smoothness_causality} appear to be automatically satisfied as a consequence of  \eqref{conds}--\eqref{inequs_centres} and \eqref{posmass}. We give details of this in the Appendix. This lends support to our conjecture that  \eqref{conds}--\eqref{inequs_centres} and \eqref{posmass} imply the smoothness and causality conditions.

To compare the solutions to the
BMPV black hole, it is useful to define the dimensionless area and
angular momentum
\begin{equation}\label{aH_eta}
  a_H \equiv \sqrt{\frac{3\sqrt{3}}{32\pi}}\frac{A_H}{Q^{3/2}}\,,\qquad 
  \eta \equiv \sqrt{6\pi\sqrt{3}}\frac{J}{Q^{3/2}}\,,
\end{equation}
where $J \equiv |J_+| = |J_-|$. For
the BMPV black hole
\begin{equation}
0\leq\eta_{\text{BMPV}}<1  \; , \qquad a_{\text{BMPV}} = \sqrt{1-\eta^2}.
\end{equation}
Therefore, a solution has the same conserved charges as BMPV iff
\eqref{equalJ} is satisfied and $0 \leq \eta <1$.  Our solutions also
carry dipole charge associated to each 2-cycle $C$ (bubble or disc),
so it is useful to also introduce a corresponding dimensionless dipole
\begin{equation}
  \label{nu_qC}
  \nu_C \equiv \sqrt{\frac{\pi}{\sqrt{3}}}\frac{|q_C|}{2Q^{1/2}}\,.
\end{equation}

We are also interested in the soliton spacetimes which have equal angular momenta. In fact, the regularity constraints \eqref{conds}, \eqref{inequs_centres}, \eqref{posmass}, \eqref{solitonreg} are compatible with the equal angular momentum condition \eqref{equalJ} only for the soliton in Figure \ref{sol:h1-11}. For this case the constraints admit a unique 1-parameter family of solutions given by
\begin{equation}
  k_2=0\,,\qquad a_1=\frac{k_1^2}{3} \,,\qquad a_2=\frac{2k_1^2}{3}, \qquad k_1\neq 0  \; .
\end{equation}
This is the soliton previously studied in~\cite{Horowitz:2017fyg}. Its physical quantities simplify substantially
\begin{equation}
\label{solitoncharges}
  Q=4\sqrt{3}\pi k_1^2\,,\qquad
  J_{\psi}=6\pi k_1^3\,,\qquad
  q_{C_1}=-q_{C_2}=-\sqrt{3}k_1\,,
\end{equation}
or in terms of the dimensionless quantities
\begin{equation}
  \eta_s=\frac{3}{2\sqrt{2}}\,,\qquad   \label{etanu_sol}
  \nu_s= \frac{1}{4}\,.
\end{equation}

\begin{figure}[t!]
\centering
\begin{tabularx}{\textwidth}{|l|X|c|c|}
  \hline
  && Horizon topology & Equal $J$?\\
  \hline\hline
  $0<a_1<a_2$ & $h_0=3$, $h_1=h_2=-1$ & $L(3,1)$ & $\times$\\
  \hline
  & $h_0=1$, $h_1=-1$, $h_2=1$  & $S^3$& $J_\phi=0$ or $J_\psi=0$ \\
  \hline
  & $h_0=1$, $h_1=1$, $h_2=-1$  & $S^3$& $J_\psi=0$ \\
  \hline
  & $h_0=-1$, $h_1=h_2=1$  & $S^3$&  $\times$\\
  \hline\hline
  $a_1<0<a_2$ & $h_1=-1$, $h_0=3$, $h_2=-1$  & $L(3,1)$& $J_\phi=0$\\
  \hline
  &  $h_1=-1$, $h_0=1$, $h_2=1$  & $S^3$&  $\times$\\
  \hline
  & $h_1=1$, $h_0=-1$, $h_2=1$  & $S^3$& $J_\phi=0$\\
  \hline
\end{tabularx}
\caption{3-centred single
  black hole solutions that allow for equal angular momenta.}
\label{table}
\end{figure}

\subsection{Spherical black hole with nontrivial topology I}%: $0<a_1<a_2$, $h_0=1$, $h_1=-1$, $h_2=1$}

Let us consider the first solution given in Figure \ref{table} which admits equal angular momenta: the
spherical black hole with $0<a_1<a_2$, $h_0=1$, $h_1=-1$,
$h_2=1$ (see Figure \ref{h=1-11_1}). This is the previously studied spherical black hole with
bubble~\cite{Kunduri:2014iga,Horowitz:2017fyg}. From \eqref{conds} it follows that its physical charges
satisfy the relation
\begin{equation}\label{Jphi_I}
  J_\phi=-\frac{1}{2}Q q_D+\frac{\pi }{\sqrt{3}}q_Bq_D(q_B- q_D)  \; ,
\end{equation}
and we can express the area of the horizon \eqref{AH} of the black hole
in terms of physical quantities,
\begin{equation}
  A_H=8\pi^2\left[
    \frac{1}{6\sqrt{3}\pi^3}\left(Q+\frac{4\pi}{\sqrt{3}}q_Bq_D\right)^3
    -
    \left(\frac{J_\psi+J_\phi}{\pi} -
      \frac{2}{\sqrt{3}}q_Dq_B^2\right)^2\right]^{1/2}\,.   \label{AreaSBHI}
\end{equation}
As can be seen from Figure \ref{table}, both $J_\phi$ and
$J_\psi$ can vanish in some subregion of the moduli space of the
solution. We will study these two cases separately.

In general, we can always solve \eqref{conds} for $\ell_0$ and $m_0$,
as $h_2 k_1 - h_1 k_2 = k_1 + k_2 \neq 0$ is guaranteed by the
constraints on the parameters. The solution is then parameterised by
the remaining 4 parameters $(k_1,k_2,a_1,a_2)$. Setting either
$J_\phi$ or $J_\psi$ to zero imposes another constraint, resulting in
a 3-parameter family of solutions in either case.

\subsubsection{$J_\phi=0$}\label{sec:zeroJphi_I}

\begin{figure}[t]
\begin{subfigure}[t]{.48\textwidth}
\includegraphics[width=\textwidth]{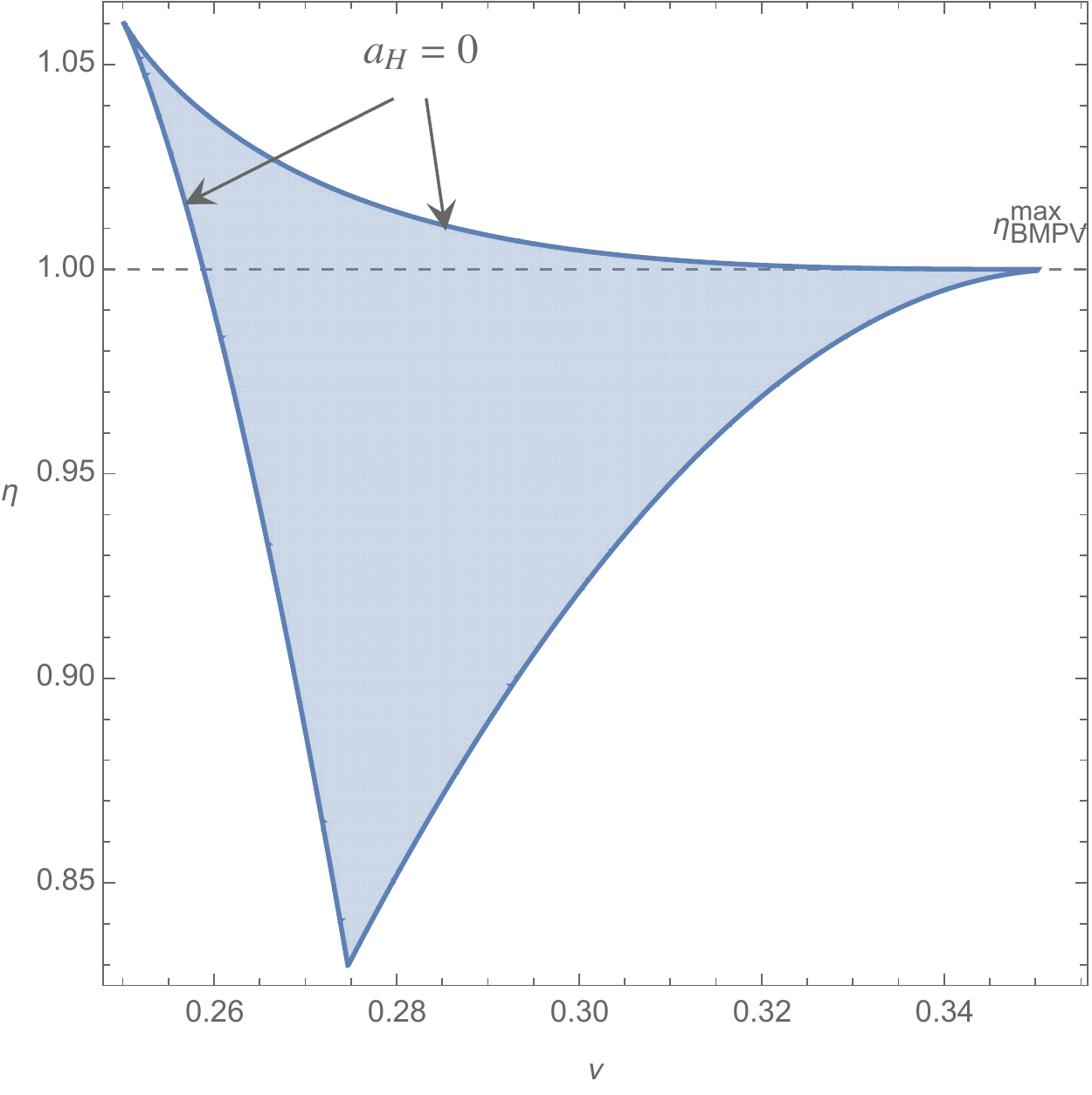}
\caption{}
\label{zeroJphi_I_region}
\end{subfigure}\hfill
\begin{subfigure}[t]{.48\textwidth}
\includegraphics[width=\textwidth]{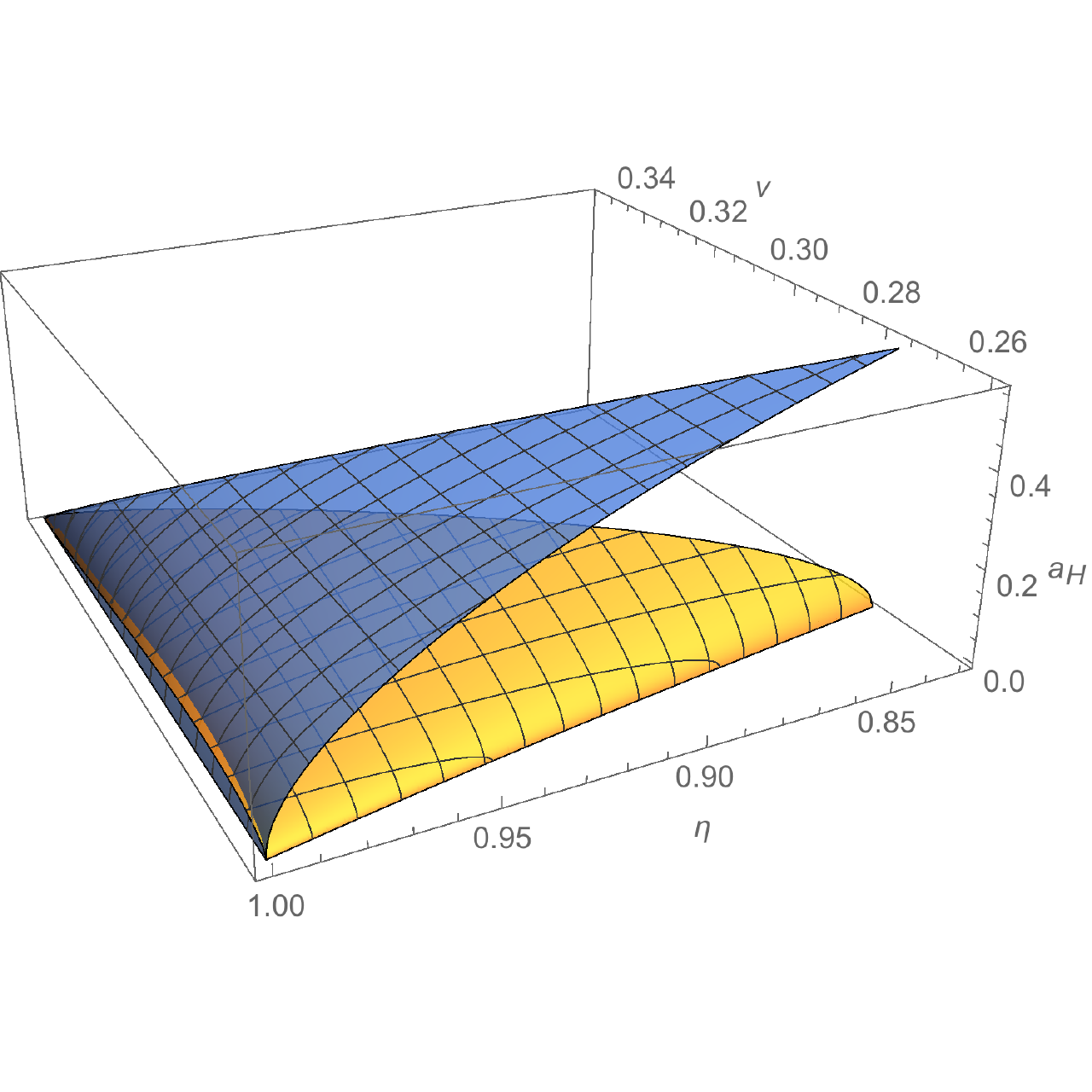}
\caption{}
\label{zeroJphi_I_3darea}
\end{subfigure}
\caption{\subref{zeroJphi_I_region} Moduli space for the
  $J_\phi=0$  spherical black hole I ($0<a_1<a_2$, $h_0=1$,
  $h_1=-1$, $h_2=1$). \subref{zeroJphi_I_3darea} Dimensionless area of the spherical black hole I (orange/lighter) and the BMPV black hole (blue/darker), in the region of overlap. Observe that $a_H>a_{\text{BMPV}}$ in a narrow region close to $\eta=1$.}
\label{zeroJphi_I}
\end{figure}

This case has been
previously studied in \cite{Horowitz:2017fyg}, which we will briefly
review here. As mentioned above, by solving \eqref{conds} for $\ell_0$
and $m_0$, one obtains a 4-parameter family of solutions, determined
by the parameters $(k_1,k_2,a_1,a_2)$. For the case at hand, we can
readily solve $J_\phi=0$ for $k_2$,
\begin{equation}
  k_2=-\frac{(2a_1-a_2)k_1}{a_1}\,,
\end{equation}
leaving a 3 parameter family solutions specified by $(a_1,a_2,k_1)$.
One can express all physical quantities in terms of the dimensionless angular momentum
$\eta$ \eqref{aH_eta} and dipole $\nu \equiv \nu_B$ \eqref{nu_qC}. In particular, the reduced area 
\begin{equation}
  a_H =
  \left[\left(16\nu^2-1\right)^3-\left(\eta+6\sqrt{2}\nu(8\nu^2-1)\right)^2
    \right]^{1/2}\,.
\end{equation}
The inequalities
\eqref{inequ_horizon}, \eqref{inequs_centres} and \eqref{posmass} 
reduce to
\begin{equation}
\begin{gathered}
 \frac{1}{4} < \nu < \frac{1}{2\sqrt{2}}\; , \qquad \text{max} \left( \eta_-(\nu), \frac{-1+ 40 \nu^2-128 \nu^4}{4 \sqrt{2} \nu} \right)< \eta < \eta_+(\nu) \,,  \label{inequs_zeroJphi_I}
\end{gathered}
\end{equation}
where 
\begin{equation}
\eta_\pm(\nu) = \pm (16\nu^2-1)^{3/2} +6 \sqrt{2}\nu (1-8\nu^2)  \; ,    \label{etapmSBH}
\end{equation}
implying the range
\begin{equation}
  \frac{-1+8\sqrt{2}}{8\sqrt{1+\sqrt{2}}} < \eta <
                                            \frac{3}{2\sqrt{2}}\; .
\end{equation}
The moduli space defined
by \eqref{inequs_zeroJphi_I} is the triangular region depicted in Figure
\ref{zeroJphi_I_region}.
Notice that the area of the horizon vanishes along two of the boundary curves defined by $\eta_\pm(\nu)$ as shown in Figure \ref{zeroJphi_I_region}.  Moreover, the point
$(\eta,\nu)=(\frac{3}{2\sqrt{2}}, \frac{1}{4})$ where those two curves
intersect corresponds to the soliton solution \eqref{etanu_sol}. In fact, expanding the black hole solution near the soliton point reveals that one can interpret the solution in this limit as a
small, nonrotating, extremal black hole 
sitting in the bubbling soliton geometry~\cite{Horowitz:2017fyg}.

Finally, let us compare the solution with the BMPV black hole. In the region of overlap and near
$\eta = 1$, \ie for high angular momentum, there is a narrow region of the
moduli space in which the black hole with
nontrivial topology has an area (hence entropy) that is greater than
that of the BMPV solution, see Figure \ref{zeroJphi_I_3darea}.

\subsubsection{$J_\psi=0$}

This case has not been previously studied.  As mentioned above, for
this general family of solutions we can always solve the constraints
for $\ell_0$ and $m_0$. We now want to also set $J_\psi=0$. This, in
general, imposes a more complicated constraint than vanishing of
$J_\phi$. Nevertheless, defining the dimensionless ratio
$\alpha=a_1/a_2$, so $0<\alpha<1$, we can solve $J_\psi=0$ (for
$a_2$), resulting in a 3-parameter set of solutions specified by
$(\alpha,k_1,k_2)$. The constraints on the parameters imply
$k_1 \neq 0$ so it is convenient to introduce the dimensionless ratio
$\kappa\equiv k_2/k_1$. In terms of these dimensionless variables we
find that the constraints on the parameters can be reduced to%
\footnote{The upper (lower) limit for $\kappa$ ($\alpha$) are
  determined by the largest real root of
  $\kappa_0^3 + 9 \kappa_0 ^2 + 18 \kappa_0 +9=0$ and
  $\alpha_0 =
  (5+3\kappa_0)/(6+6\kappa_0+3\kappa_0^2+\kappa_0^3)$.}
\begin{equation}\label{inequs_zeroJpsi_I}
  \begin{gathered}
    -1<\kappa <-0.773318\, ,  \qquad a_H^2>0\,,\qquad
    \frac{5+3 \kappa}{6+6\kappa+3\kappa^2+\kappa^3}
    <\alpha <
    \alpha_{+}(\kappa)\,,
  \end{gathered}
\end{equation}
where
\begin{equation}
  \alpha_{+}(\kappa) = \frac{
    - 8 - 7\kappa + \kappa ^3 
    +(1+\kappa) \sqrt{64 + 152\kappa + 149\kappa^2 + 70\kappa^3 + 13\kappa^4}}
  {2 \kappa  (2+\kappa)(3+3\kappa+\kappa^2)}\,,
\end{equation}
the reduced area $a_H$ is a complicated function of $\alpha, \kappa$, and the resulting
range of $\alpha$ is
\begin{equation}
  0.995673<\alpha<1\,.
\end{equation}

To translate this to physical parameters, let us again introduce a dimensionless dipole $\nu \equiv \nu_D$
(note that here we chose $\nu$ to be proportional to the dipole charge
$q_D$ rather than $q_B$). In terms of $\alpha$ and $\kappa$, the
dimensionless charges $\eta$ and $\nu$ are given by
\begin{align}
  \label{eta_zeroJpsi_I}
  \eta &=
  \frac{3\sqrt{3(2+\kappa)}(1-\alpha)(1+\kappa)\left(-1+\alpha(2+\kappa)\right)\left(-5-3
  \kappa + \alpha  (\kappa ^3+3 \kappa ^2+6 \kappa
  +6)\right)}
  {2\left[
  7  +9 \kappa + 3 \kappa^2 - \alpha
  (8+7\kappa-\kappa^3)-\alpha^2\kappa(6+9\kappa+5\kappa^2+\kappa^3)
  \right]^{3/2}}\,,\\
  \label{nu_zeroJpsi_I}
 \nu &=
       \frac{\sqrt{3(2+\kappa)}(1-\alpha)(1+\kappa)}
  {2\sqrt{2}\left[
  7  +9 \kappa + 3 \kappa^2 - \alpha
  (8+7\kappa-\kappa^3)-\alpha^2\kappa(6+9\kappa+5\kappa^2+\kappa^3)
  \right]^{1/2}}\,, 
\end{align}
where positivity of each factor in the numerators and denominators is guaranteed by \eqref{inequs_zeroJpsi_I}.
In fact, within the region of interest \eqref{inequs_zeroJpsi_I}, we
can uniquely invert \eqref{eta_zeroJpsi_I} and \eqref{nu_zeroJpsi_I},
giving 
\begin{equation}
  \kappa =
  -1 + 4\nu^2
  \left(-2\nu^2+\sqrt{2\nu^2(1+2\nu^2)+\tfrac{\sqrt{2}}{3}\eta\nu}\right)^{-1}
\end{equation}
and some complicated expression for $\alpha$.  To derive this inverse we used \eqref{Jphi_I} to solve for $q_B$ in terms of the other physical variables.

\begin{figure}[t]
\begin{subfigure}[t]{.48\textwidth}
\includegraphics[width=\textwidth]{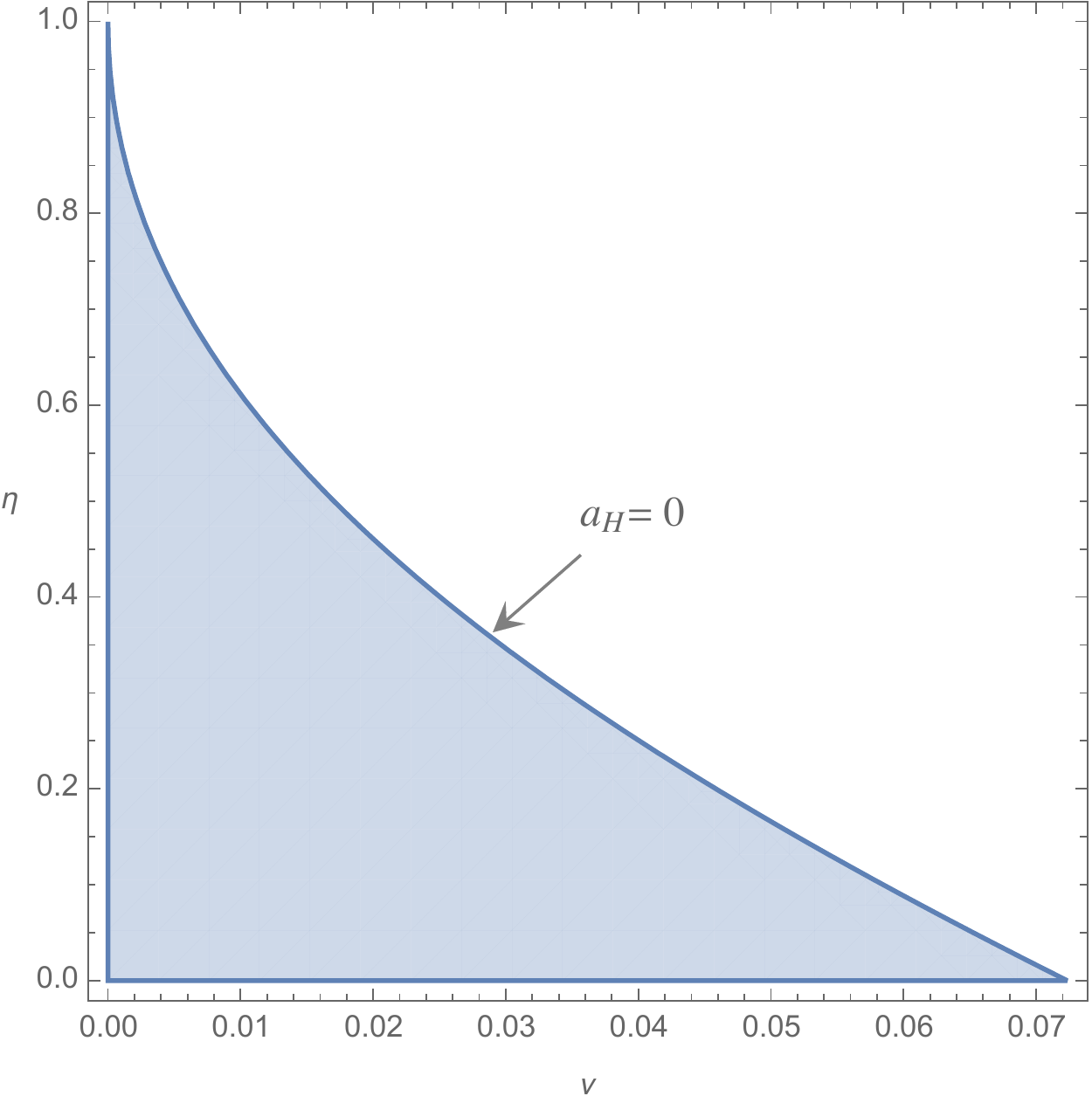}
\caption{}
\label{zeroJpsi_I_region}
\end{subfigure}\hfill
\begin{subfigure}[t]{.48\textwidth}
\includegraphics[width=\textwidth]{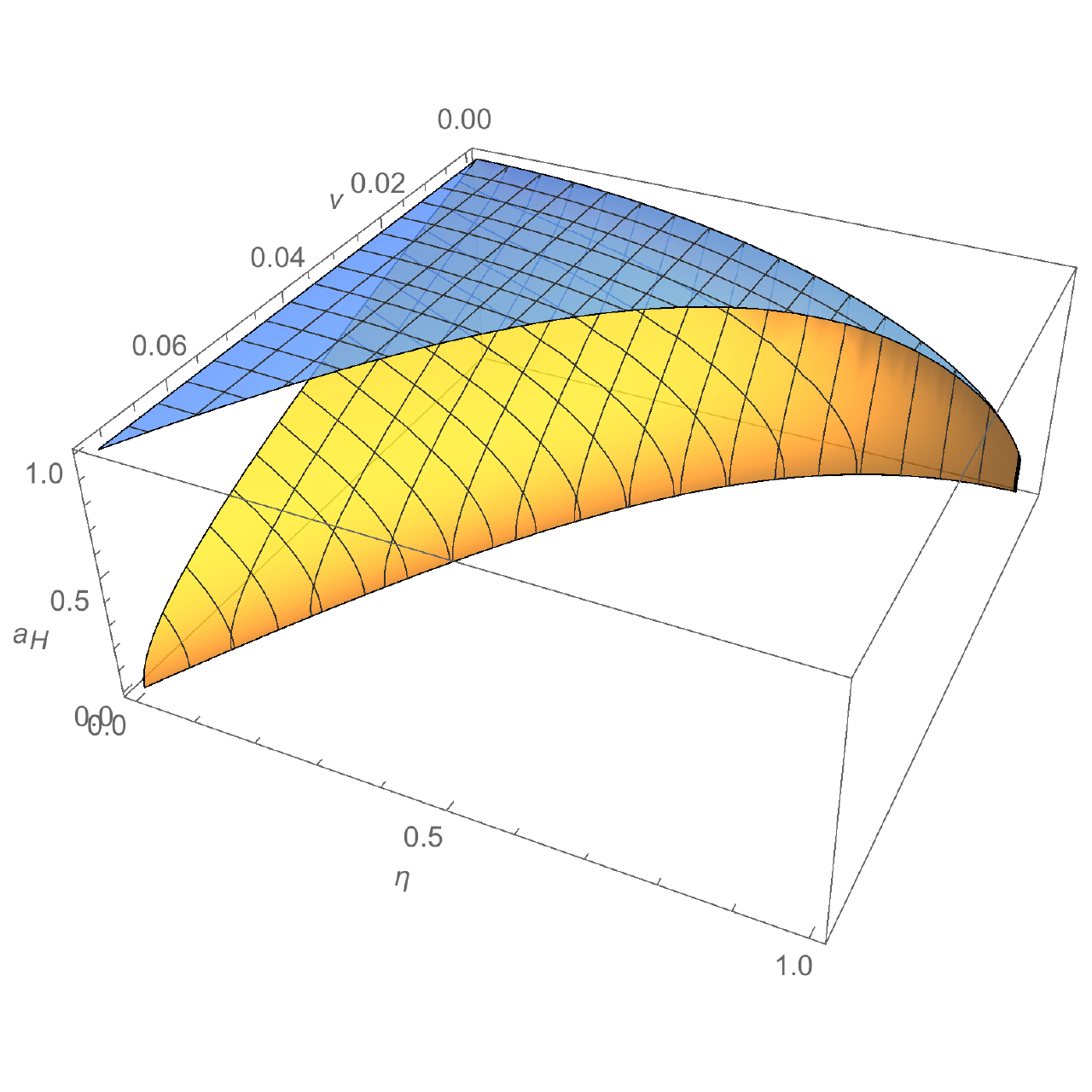}
\caption{}
\label{zeroJpsi_I_3darea}
\end{subfigure}
\caption{\subref{zeroJphi_I_region} Moduli space of the
  $J_\psi=0$ spherical black hole I ($0<a_1<a_2$, $h_0=1$,
  $h_1=-1$, $h_2=1$). \subref{zeroJphi_I_3darea} Dimensionless area (orange/lighter) as compared to that of the BMPV
  black hole (blue/darker) in the region of overlap. In this case $a_H<a_\text{BMPV}$ throughout the overlap region.}
\label{zeroJpsi_I}
\end{figure}

This also allows us to write the
area in terms of $\eta$ and $\nu$ as 
\begin{multline}
  a_H =
  \Bigg[\left(1+8\nu^2 -
      4\sqrt{2\nu^2(1+2\nu^2)+\tfrac{\sqrt{2}}{3}\eta\nu}\right)^3\\
    - \left(\eta + 6\sqrt{2}\nu\left(
        1+4\nu^2 - 2\sqrt{2\nu^2(1+2\nu^2)+\tfrac{\sqrt{2}}{3}\eta\nu}
        \right)\right)^2\Bigg]^{1/2}\,,
\end{multline}
and the region \eqref{inequs_zeroJpsi_I} in terms of the physical parameters reduces to 
%given by
\begin{equation}\label{inequs_zeroJpsi_I_EtaNu}
  \eta >0\,,\qquad \nu >0\,,\qquad a_H^2>0\,,
\end{equation}
leading to the ranges
\begin{equation}
  0<\eta<1\,, \qquad 0< \nu < 0.072361 \; .  
\end{equation}
The exact upper bound of $\nu$ is the unique positive root of $a_H(\eta=0)=0$. The resulting moduli space is the region depicted in Figure
\ref{zeroJpsi_I_region}.

The upper bound for $\eta$ at fixed $\nu$ (or vice-versa) corresponds to $a_H=0$. The lower bound $\eta=0$ 
corresponds to the curve 
\begin{equation}
-5-3 \kappa + \alpha \left(\kappa ^3+3 \kappa ^2+6 \kappa +6\right)=0.  \label{bdry_lowerSBHI}
\end{equation}
The lower bound $\nu=0$ for $\eta>0$ has no well defined meaning  in
terms of $\alpha$ and $\kappa$, as a consequence of the form of
\eqref{eta_zeroJpsi_I} and \eqref{nu_zeroJpsi_I}. The corner
$(\eta,\nu)=(0,0)$ corresponds to the limit $\alpha\to 1$,
$\kappa\to-1$ (or equivalently in terms of the original parameters
$a_1\to a_2$, $k_1\to-k_2$). This can be seen as follows. 

Consider the limit
$(\alpha,\kappa) \to (1,-1)$ along a curve within the region defined
by \eqref{inequs_zeroJpsi_I}. The limiting point $(\alpha,\kappa) = (1,-1)$ corresponds to the corner of the moduli space defined by the intersection of two boundary curves \eqref{bdry_lowerSBHI} and $a_H=0$. We can expand each curve about this corner as a series in $(\kappa+1)$:  for \eqref{bdry_lowerSBHI} we find
\begin{equation}
\alpha =
 1-\frac{1}{2}(\kappa+1)^3+\frac{3}{4}(\kappa+1)^4
  -\frac{9}{8}(\kappa+1)^5 + \Ord((\kappa+1)^6)\, ,
  \end{equation}
  whereas for $a_H=0$ we find
  \begin{equation}
 \alpha =
  1-\frac{1}{2}(\kappa+1)^3+\frac{3}{4}(\kappa+1)^4
  -\frac{3}{4}(\kappa+1)^5 + \Ord((\kappa+1)^6)\,.
  \end{equation}
Note that they agree up to fourth order in $(\kappa+1)$! Thus any
smooth curve approaching the point $(1,-1)$ from within the moduli space
will be of the form
\begin{equation}
  \alpha =
  1-\frac{1}{2}(\kappa+1)^3+\frac{3}{4}(\kappa+1)^4
  + \alpha^{(5)}(\kappa+1)^5 + \Ord((\kappa+1)^6)
\end{equation}
for some
\begin{equation}
  -\frac{9}{8}<\alpha^{(5)}<-\frac{3}{4}\,.
\end{equation}
Approaching the corner along such a curve, we find the physical charges
\begin{equation}
  Q \to -\frac{4}{\sqrt{3}}(3+4\alpha^{(5)}) k_1^2 \pi\,,\qquad
  J_\phi \to 0\,,\qquad
  q_D \to 0 \,, \qquad
  q_B\to -\sqrt{3} k_1\,,
\end{equation}
and the area
\begin{equation}
  A_H \to \frac{32\sqrt{2}}{3\sqrt{3}} \sqrt{-(3+4\alpha^{(5)})^3} |k_1|^3 \pi^2\,.
\end{equation}
Hence the dimensionless quantities tend to
\begin{equation}
  a_H  \to 1\,,
  \qquad \eta\to 0\,,\qquad \nu\to 0\,.
\end{equation}
These are the physical quantities of a Reissner--Nordstr\"om solution. In terms of our solution this limit may be understood as arising from
an effective ``cancelling out'' of two centres (recall $a_1 \to a_2, k_1\to - k_2$ in this limit). We may
thus interpret the solution near this limit as a Reissner--Nordstr\"om
solution with a bubble added in the exterior of the black hole. This
is in contrast to the solution with $J_\phi=0$, which
was interpreted as a black hole sitting in a soliton spacetime.\footnote{It is of course still possible that this solution has a soliton limit in the more general moduli space of solutions with unequal angular momenta.}

For the present class of solutions $0<\eta<1$, so the entire parameter space \eqref{inequs_zeroJpsi_I_EtaNu}
overlaps with the BMPV solution. Furthermore, in contrast to the previous case, this shows that there are solutions with the same conserved charges as any rotating BMPV black hole (i.e. $\eta >0$), no matter how small $\eta$!
Finally, one can show that
\begin{equation}
  a_H<a_{\text{BMPV}}
\end{equation}
throughout the moduli space \eqref{inequs_zeroJpsi_I_EtaNu}. This is depicted in Figure \ref{zeroJpsi_I_3darea}. Thus for this class of solutions the entropy is always subdominant to the BMPV black hole.

\subsection{Spherical black hole with nontrivial topology II}

We will now study the second solution in Figure \ref{table} which admits equal angular momentum: the spherical black hole with $0<a_1<a_2$, $h_0=1$,
  $h_1=1$, $h_2=-1$ (see Figure \ref{h=11-1}).  This solution has not been studied before. The
charges of the solution obey the constraint
\begin{equation}
  J_\phi=-\frac{1}{2}Q q_D+\frac{\pi}{3\sqrt{3}}q_D(3q_B^2
  - 3q_Bq_D+2 q_D^2)\,,   \label{constraintSBHII}
\end{equation}
and the area of the horizon as a function of the charges is given by
\begin{multline}
  A_H=8\pi^2\Bigg[
    \frac{1}{6\sqrt{3}\pi^3}\left(Q+\frac{4\pi}{\sqrt{3}}q_D(q_D-q_B)\right)^3\\
    -
    \left(\frac{J_\psi+J_\phi}{\pi} -
      \frac{2}{3\sqrt{3}}q_D(3q_B^2-6q_Bq_D+4q_D^2)\right)^2\Bigg]^{1/2}\,.
\end{multline}
As shown in Figure \ref{table}, this solution admits equal angular momentum only if $J_\psi=0$. 

The analysis here is very similar to
the $J_\psi=0$ case of spherical black hole I.  As in that case, we can always solve \eqref{conds}
for $\ell_0$, $m_0$, since $h_2k_1-h_1k_2=-k_1-k_2\neq 0$ by the constraints on the parameters. Now, imposing $J_\psi=0$, we obtain a
three-parameter family of solutions parameterised by
$(\alpha, k_1,k_2)$ where again $\alpha = a_1/a_2$ (hence by definition
$0<\alpha<1$).  As in the previous case we may introduce $\kappa
\equiv k_2/k_1$ since the constraints on the parameters guarantee
$k_1\neq 0$. The resulting set of constraints on the parameters can be
reduced to\footnote{The exact lower limits for $\alpha$ and $\kappa$ are given by the smallest real root of
  $19 + 150 \kappa_0 + 537 \kappa_0^2 + 1163 \kappa_0^3 + 1590
  \kappa_0^4 + 1284 \kappa_0^5 + 424 \kappa_0^6 =0$ and
  $\alpha_0= (1+5\kappa_0+10\kappa_0^2+8\kappa_0^3
  )/(2+8\kappa_0+13\kappa_0^2+9\kappa_0^3)$.}
\begin{equation}\label{inequs_zeroJpsi_II}
  \begin{gathered}
    -1.307191<\kappa <-1\, ,  \qquad a_H^2>0\,,\qquad
    \alpha_{-}(\kappa)
    <\alpha <
    \frac{1+5\kappa+10\kappa^2+8\kappa^3}{2+8\kappa+13\kappa^2+9\kappa^3}\,,
  \end{gathered}
\end{equation}
where
\begin{multline}
\alpha_{-}(\kappa) = \frac{1}{2\kappa^2 (3 + 9\kappa + 7\kappa^2)}\Bigg[(1+ 2\kappa) \left(
        -2-5\kappa-2\kappa^2+3\kappa^3\right) \\
      - (1+\kappa)\sqrt{
      (1+2\kappa)(2+3\kappa)(2+7\kappa+10\kappa^2+7\kappa^3+6\kappa^4)}
    \Bigg]\,,
\end{multline}
$a_H$ is a complicated function of $\alpha, \kappa$, and the resulting
range of $\alpha$ is
\begin{equation}
  0.995433<\alpha<1\,.
\end{equation}
Defining $\nu \equiv \nu_D$ as in \eqref{nu_qC}, the
dimensionless angular momentum and dipole are given by
\begin{align}
  \label{eta_II}
  \eta &= 3 \sqrt{3\kappa(1+2\kappa)} (\alpha-1) (1+\kappa)
         \nonumber\\
  & \phantom{= }\times \frac{(1+2\kappa-\alpha\kappa)\left(1 + 5\kappa + 10\kappa^2 + 8 \kappa
         ^3 -\alpha
         \left(2+8\kappa+13\kappa^2+9\kappa^3\right)\right)}
    {2 \left[ -1-5 \kappa-9 \kappa ^2-6 \kappa ^3
         +\alpha  \left(2+9 \kappa+12 \kappa ^2+\kappa ^3-6 \kappa ^4
         \right)
         +\alpha ^2 \left(3+9 \kappa+ 7\kappa ^2 \right) \kappa ^2
         \right]^{3/2}}\,,\\
  \label{nu_II}
  \nu &= \frac{ \sqrt{3\kappa(1+2\kappa)} (\alpha-1) (1+\kappa)
        }
         {2\sqrt{2} \left[ -1-5 \kappa-9 \kappa ^2-6 \kappa ^3
         +\alpha  \left(2+9 \kappa+12 \kappa ^2+\kappa ^3-6 \kappa ^4
         \right)
         +\alpha ^2 \left(3+9 \kappa+ 7\kappa ^2 \right) \kappa ^2
         \right]^{1/2}}\,,
\end{align}
where positivity of the numerators and denominators follows from the above inequalities. In the region of interest defined by
\eqref{inequs_zeroJpsi_II}, we may invert \eqref{eta_II} and \eqref{nu_II} to obtain
\begin{equation}
  \kappa = \frac{-6\nu^2 + \sqrt{3}
    \sqrt{2\nu^2(3-10\nu^2)+\sqrt{2}\eta\nu}}{18 \nu ^2-\sqrt{3}
    \sqrt{2\nu^2(3-10\nu^2)+\sqrt{2}\eta\nu}}
\end{equation}
and (again) some complicated expression for $\alpha$. To derive this inverse we used \eqref{constraintSBHII} to solve for $q_B$ in terms of the other physical quantities. 

\begin{figure}[t]
\begin{subfigure}[t]{.48\textwidth}
\includegraphics[width=\textwidth]{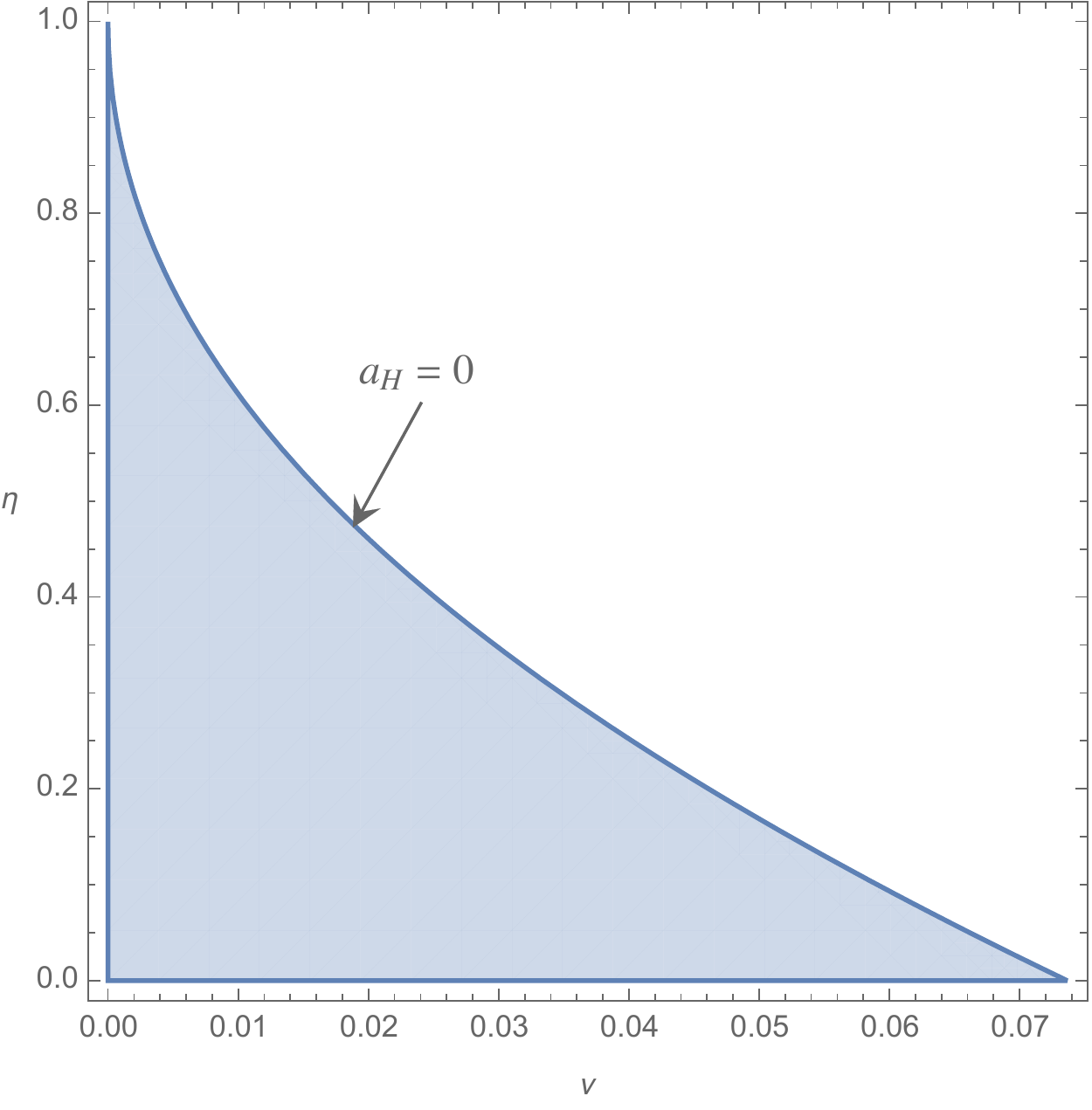}
\caption{}
\label{zeroJpsi_II_region}
\end{subfigure}\hfill
\begin{subfigure}[t]{.48\textwidth}
\includegraphics[width=\textwidth]{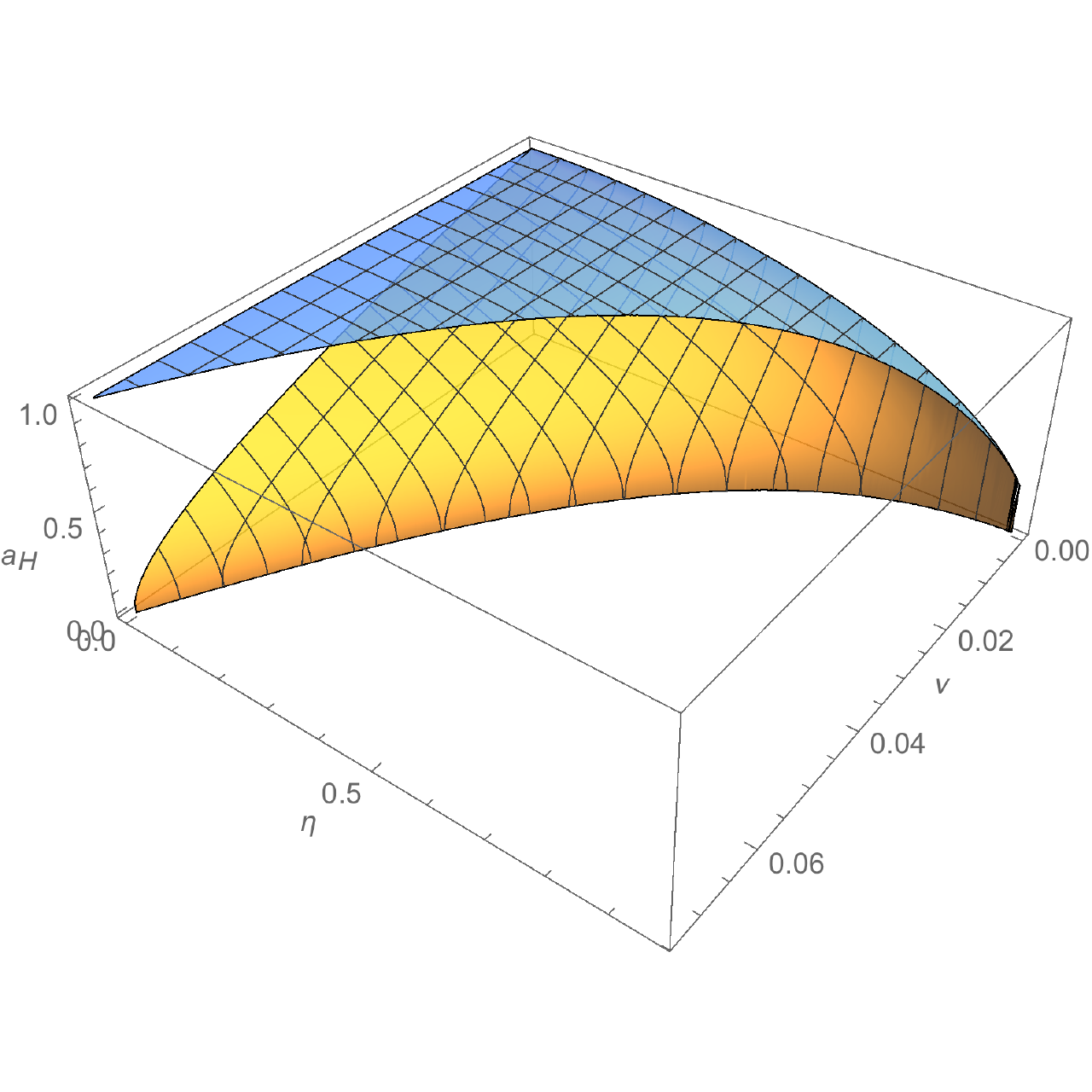}
\caption{}
\label{zeroJpsi_II_3darea}
\end{subfigure}
\caption{\subref{zeroJpsi_II_region} Moduli space for the $J_\psi=0$ 
  spherical black hole II ($0<a_1<a_2$, $h_0=1$, $h_1=1$, $h_2=-1$). \subref{zeroJpsi_II_3darea} Dimensionless area of the
  horizon (orange/lighter) as compared to that of the BMPV black hole (blue/darker) in the region of overlap. In this case $a_H<a_\text{BMPV}$ throughout the overlap region.}
\label{zeroJpsi_II}
\end{figure}

This also allows us to write the dimensionless area of the horizon as
\begin{multline}
  a_H =
  \Bigg[\left(1+8\nu^2 -
      \frac{4}{\sqrt{3}}\sqrt{2\nu^2(3-10\nu^2)+\sqrt{2}\eta\nu}\right)^3\\
    - \left(\eta + 2\sqrt{2}\nu\left(
        3+4\nu^2 - 2\sqrt{3}\sqrt{2\nu^2(3-10\nu^2)+\sqrt{2}\eta\nu}
        \right)\right)^2\Bigg]^{1/2}\, ,
\end{multline}
and the moduli space
\eqref{inequs_zeroJpsi_II} in terms of the physical variables is now given  by
\begin{equation}\label{inequs_zeroJpsi_II_EtaNu}
  \eta >0\,,\qquad \nu >0\,,\qquad a_H^2>0\, .
\end{equation}
This implies the ranges
\begin{equation}
  0<\eta<1\,, \qquad 0< \nu < 0.073674\,,
\end{equation}
where the upper bound for $\nu$ is given by the positive root of $a_H(\eta=0)=0$. The resulting moduli space is the region depicted in Figure \ref{zeroJpsi_II_region}.

The upper bound for $\eta$ at fixed $\nu$ corresponds to $a_H=0$, whereas the lower bound for $\eta$ corresponds to the curve
\begin{equation}
1+5\kappa+10\kappa^2+8\kappa^3
-\alpha(2+8\kappa+13\kappa^2+9\kappa^3)= 0  \; .   \label{bdry_lowerSBHII}
\end{equation}
The corner $(\eta,\nu)=(0,0)$ corresponds to the limit
$\alpha\to 1$, $\kappa\to-1$ (or $a_1\to a_2$, $k_1\to-k_2$) where the two boundary curves \eqref{bdry_lowerSBHII} and $a_H=0$ intersect.  As in the previous case,  we may expand any curve that approaches this limit from inside the moduli space to find
\begin{equation}
  \alpha =
  1-\frac{1}{2}(\kappa+1)^3+\frac{9}{4}(\kappa+1)^4
  + \alpha^{(5)}(\kappa+1)^5 + \Ord((\kappa+1)^6)
\end{equation}
for some
\begin{equation}
  \frac{53}{8}<\alpha^{(5)}< 7 \,, 
\end{equation}
where the lower bound corresponds to \eqref{bdry_lowerSBHII} and the upper bound to $a_H=0$. We then find the physical charges along such a curve are
\begin{equation}
  Q \to \frac{16}{\sqrt{3}}(7-\alpha^{(5)}) k_1^2 \pi\,,\qquad
  J_\phi \to 0\,,\qquad
  q_D \to 0 \,, \qquad
  q_B\to -\sqrt{3} k_1\,,
\end{equation}
and the area
\begin{equation}
  A_H \to \frac{256\sqrt{2}}{3\sqrt{3}} \sqrt{(7-\alpha^{(5)})^3} |k_1|^3 \pi^2\,.
\end{equation}
Therefore, in this limit 
\begin{equation}
  a_H  \to 1\,,
  \qquad \eta\to 0\,,\qquad \nu\to 0\,,
\end{equation}
just as we found in the previous case. Thus we also interpret this solution in this limit as a Reissner--Nordstr\"om black hole with a bubble added to the exterior.

Again, this family of solutions has the same charges as any BMPV black hole with $\eta >0$. Comparing the horizon area of the solution to that of the BMPV black
hole again reveals that 
\begin{equation}
a_H<a_{\text{BMPV}}
\end{equation}
throughout the moduli space,
see Figure \ref{zeroJpsi_II_3darea}, so this solution is also entropically subdominant to the BMPV black hole.  

\subsection{Black lens}

We now study the third solution in Figure \ref{table} which admits equal angular momenta: the $L(3,1)$ black lens defined by $a_1<0<a_2$,
  $h_1=-1$, $h_0=3$, $h_2=-1$ (see Figure \ref{h=-13-1}). This solution has not been previously studied.  The physical charges obey the constraint
  \begin{equation}
  J_\phi= -\frac{1}{2} Q(q_{D_1}+q_{D_2}) + \frac{\pi}{3 \sqrt{3}} \left( q_{D_1}^3 + q_{D_2}^3 + (q_{D_1}+ q_{D_2})^3 \right)  \; ,
  \end{equation}
  and the area as a function of the charges is
  \begin{multline}
  A_H=  8 \pi^2 \left[ 
  \frac{1}{2 \sqrt{3} \pi^3} \left( Q-\frac{4 \pi}{3 \sqrt{3}}  \left( q_{D_1}^2+q_{D_1}q_{D_2} +q_{D_2}^2\right) \right)^3  \right. \\  \left.-  \left( \frac{3 J_\psi }{\pi}+\frac{Q (q_{D_1}-q_{D_2}) }{2 \pi } + \frac{(q_{D_1}-q_{D_2}) (2q_{D_1}+q_{D_2})(q_{D_1}+ 2 q_{D_{2}})}{9 \sqrt{3}} \right)^2   \right]^{1/2}  \; .
  \end{multline}
  As shown in  Figure \ref{table} this solution admits equal angular momenta only if $J_\phi=0$. We will now study this special case in detail.
  
  Solving the constraints \eqref{conds} together with the equal angular momentum condition $J_\phi=0$ gives,
\begin{equation}
  k_2=k_1\,, \qquad a_1=-a_2\,,\qquad m_0 = \frac{k_1}{2}(9a_2-3k_1^2+3\ell_0) \; .
\end{equation}
The solution now depends on the remaining three parameters $(a_2,k_1,\ell_0)$.
The inequalities \eqref{inequ_horizon}, \eqref{inequs_centres} then reduce to
\begin{equation}
  3k_1^2-\ell_0 -a_2>0\,,\qquad 3 \ell_0^3-\frac{9}{4}k_1^2 \left(9 a_2-3 k_1^2+3 \ell_0\right)^2>0\,,   \label{lens:inequs}
\end{equation}
which also guarantee \eqref{posmass} is obeyed.  

\begin{figure}[t]
\begin{subfigure}[t]{.48\textwidth}
\includegraphics[width=\textwidth]{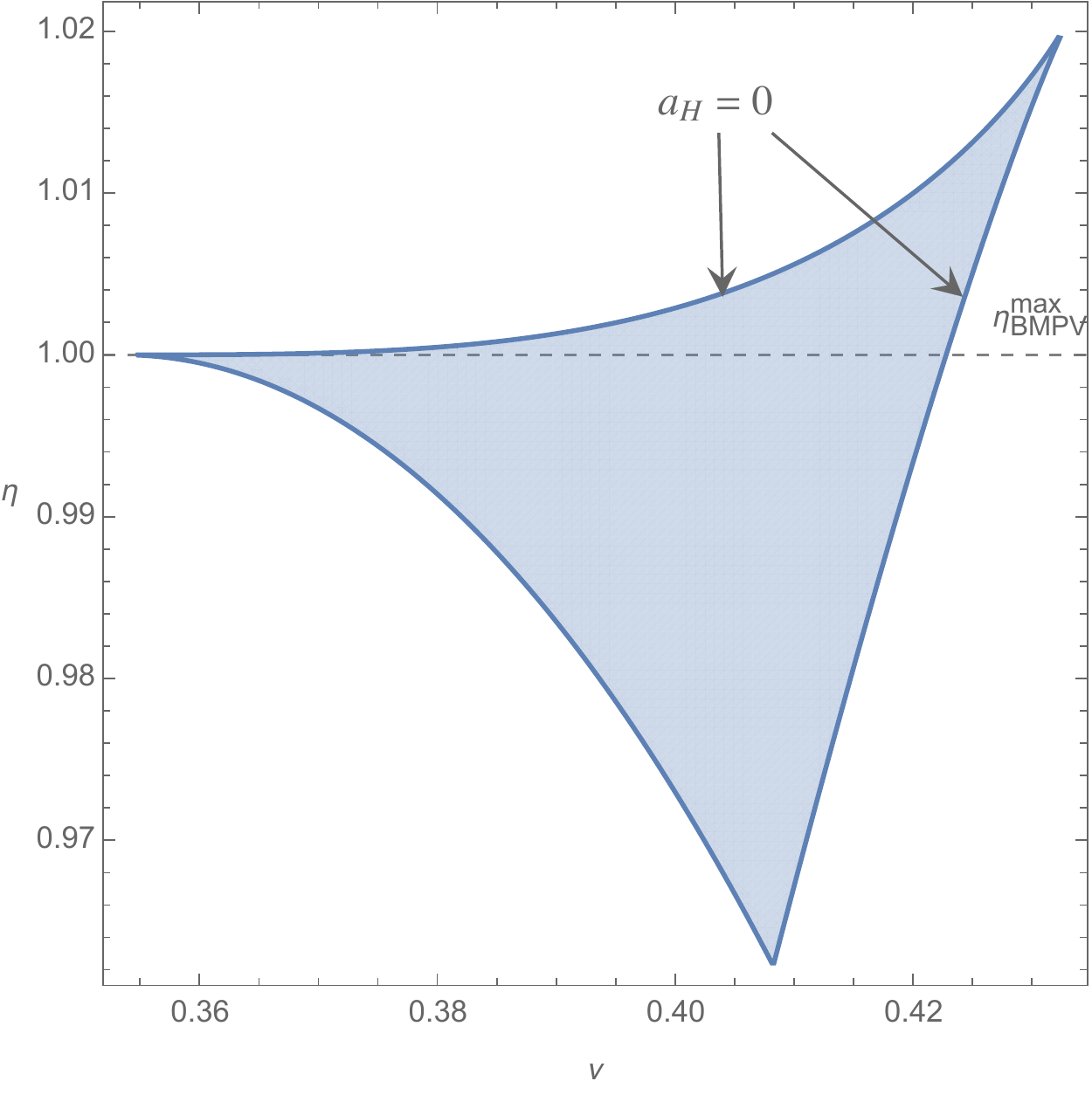}
\caption{}
\label{zeroJphi_lens_region}
\end{subfigure}\hfill
\begin{subfigure}[t]{.48\textwidth}
\includegraphics[width=\textwidth]{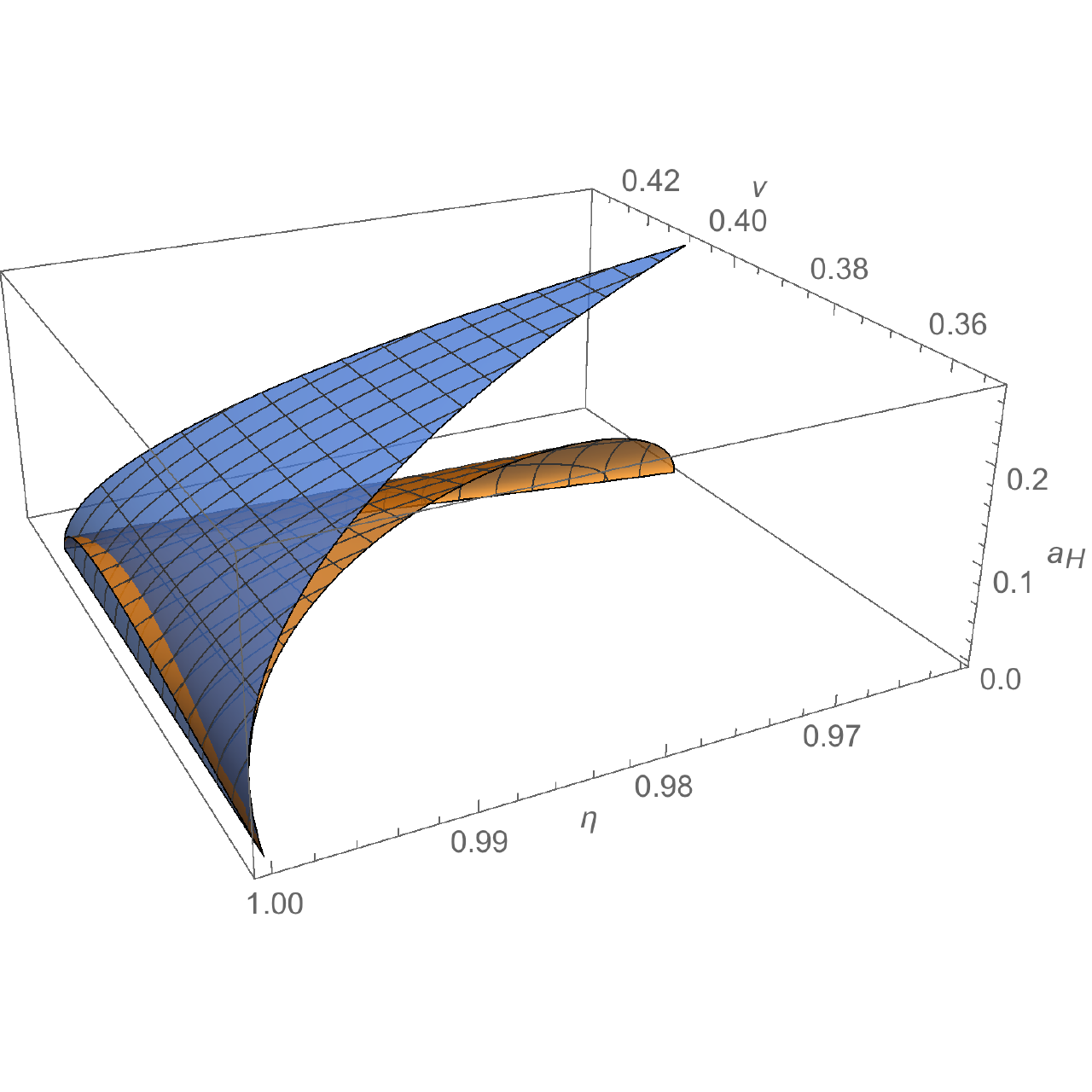}
\caption{}
\label{zeroJphi_lens_3darea}
\end{subfigure}
\caption{\subref{zeroJphi_lens_region} Moduli space for the $L(3,1)$ black lens  ($a_1<0<a_2$, $h_1=-1$, $h_0=3$, $h_2=-1$), with $J_\phi=0$. \subref{zeroJphi_lens_3darea} Dimensionless area
  of the black lens  (orange/lighter) and of the BMPV black hole
  (blue/darker), within the region of overlap. Observe that $a_H>a_{\text{BMPV}}$ in a narrow region close to $\eta=1$. }
\label{zeroJphi_lens}
\end{figure}

The dimensionless angular momentum \eqref{aH_eta} and dipole $\nu\equiv \nu_{D_1}$ \eqref{nu_qC} are 
\begin{equation}
\eta = \frac{ 9 |k_1| ( \ell_0+ 3k_1^2+a_2)}{2 ( \ell_0+ 6 k_1^2)^{3/2}}, \qquad \nu = \frac{3 |k_1|}{2 \sqrt{2} \sqrt{ \ell_0+ 6 k_1^2}} \; ,
\end{equation}
where positivity of both the numerators and denominators follows from the above inequalities \eqref{lens:inequs}. One can invert these to obtain
\begin{equation}
\frac{k_1^2}{a_2} = \frac{16 \nu^3}{3( \sqrt{2} \eta - 6 \nu+ 16 \nu^3)} \; ,\qquad \frac{\ell_0}{a_2} =  \frac{2\nu(3 -16 \nu^2)}{\sqrt{2} \eta - 6 \nu+ 16 \nu^3}  \; ,
\end{equation}
where the denominator is positive as a consequence of  the inequalities.  We may now express the moduli space defined by  \eqref{lens:inequs} in terms of the physical variables. We find this reduces to
\begin{equation}
\begin{gathered}
 \frac{1}{2\sqrt{2}}< \nu  < \frac{\sqrt{3}}{4} \,,\qquad \text{max}\left(\eta_-(\nu), \sqrt{2}(3 \nu- 8 \nu^3 ) \right) <\eta < \eta_+(\nu)\,,
\end{gathered}
\end{equation}
\begin{equation}
\eta_\pm(\nu) = \pm \tfrac{1}{\sqrt{3}} \left(1 -\tfrac{16}{3} \nu ^2\right)^{3/2}+ \tfrac{4}{9\sqrt{2}} \nu  \left(9 - 8 \nu
        ^2\right)
\end{equation}
implying the range
\begin{equation}
 \frac{5}{3\sqrt{3}}< \eta < \frac{5}{2\sqrt{6}} \; .
\end{equation}
The resulting moduli space defined by this is the triangular region depicted in Figure \ref{zeroJphi_lens_region}.
The reduced area  is given by
\begin{equation}
  a_H = \left[
    3 \left(1 -\tfrac{16}{3} \nu ^2\right)^3
    - \left(3 \eta -\tfrac{4}{3\sqrt{2}} \nu  \left(9 - 8 \nu
        ^2\right)\right)^2
  \right]^{1/2}\,.
\end{equation}
As in the case with $J_\phi=0$ discussed in section
\ref{sec:zeroJphi_I}, the region is bounded by three curves, along two
of which  $a_H$ vanishes (corresponding to the bounds $\eta_\pm(\nu)$).
Similarly, the region 
extends beyond the BMPV upper bound $\eta=1$. Furthermore, close to this bound we find a region in which the black lens has higher entropy
than the BMPV black hole. The areas of the BMPV black hole and black lens are plotted in Figure \ref{zeroJphi_lens_3darea}.

It should be noted that in contrast to the spherical black hole I discussed in section \ref{sec:zeroJphi_I}, there is no possibility of a soliton limit of the black lens solution (the soliton requires $h_0=\pm 1$).

\subsection{Spherical black hole with nontrivial topology III}

Let us now consider the fourth and final solution in Figure \ref{table} with equal angular momenta: the spherical black hole with $a_1<0<a_2$, $h_1=1$, $h_0=-1$, $h_2=1$ (see Figure \ref{h=1-11_2}). Again, this solution has not been previously analysed. The physical charges obey the constraint
  \begin{equation}
  J_\phi= -\frac{1}{2} Q( q_{D_1}+ q_{D_2}) - \frac{\pi}{\sqrt{3}} q_{D_1} q_{D_2} (q_{D_1}+q_{D_2})
  \end{equation}
  and the area as a function of the charges is
  \begin{equation}
  A_H=    \left[  -\frac{1}{6 \sqrt{3} \pi^3} \left( Q + \frac{4\pi}{\sqrt{3}} q_{D_1} q_{D_2} \right)^3 - \left( \frac{J_{\psi}}{\pi} + \frac{ Q( q_{D_2}- q_{D_1})}{2\pi} + \frac{q_{D_1} q_{D_2} (q_{D_2}- q_{D_1}) }{\sqrt{3}} \right)^2 \right]^{1/2}  \; .
  \label{AreaSBHIII}
  \end{equation}
  As shown in Figure \ref{table} this solution admits equal angular momenta only if $J_\phi=0$.  The analysis of this solution very similar to that of the black
lens solutions described in the previous section. 

Solving the constraints
\eqref{conds} together with the equal angular momentum condition $J_\phi=0$ gives
\begin{equation}
  k_2=k_1\,, \qquad a_1=-a_2\,,\qquad m_0 = \frac{k_1}{2}(3a_2-k_1^2-3\ell_0)
\end{equation}
so that the solution is again described by three parameters, $(a_2,k_1,\ell_0)$.
The inequalities \eqref{inequ_horizon}, \eqref{inequs_centres}, \eqref{posmass} constraining the parameter space then reduce to
\begin{equation}\label{inequs_zeroJphi_III}
  k_1^2+\ell_0 +a_2>0\,,\qquad  -\ell_0^3-\frac{1}{4}k_1^2 \left(-3
    a_2 + k_1^2 + 3 \ell_0\right)^2>0\,,\qquad 2k_1^2+\ell_0> 0.
\end{equation}
The dimensionless angular momentum \eqref{aH_eta} and dipole $\nu\equiv \nu_{D_1}$ \eqref{nu_qC} are now
\begin{equation}
\eta = \frac{ |k_1| ( 3\ell_0+ 5 k_1^2+3 a_2)}{2 ( \ell_0+ 2 k_1^2)^{3/2}}, \qquad \nu = \frac{|k_1|}{2 \sqrt{2} \sqrt{ \ell_0+ 2 k_1^2}} \; ,
\end{equation}
where positivity of the numerators follows from the above inequalities \eqref{inequs_zeroJphi_III}.  Inverting these we obtain
\begin{equation}
\frac{k_1^2}{a_2} = \frac{48 \nu^3}{( \sqrt{2} \eta - 6 \nu+ 16 \nu^3)} \; ,\qquad \frac{\ell_0}{a_2} =  \frac{6\nu(1 -16 \nu^2)}{\sqrt{2} \eta - 6 \nu+ 16 \nu^3}  \; ,
\end{equation}
where the denominator is positive as a consequence of  the inequalities.  

\begin{figure}[t]
\begin{subfigure}[t]{.48\textwidth}
\includegraphics[width=\textwidth]{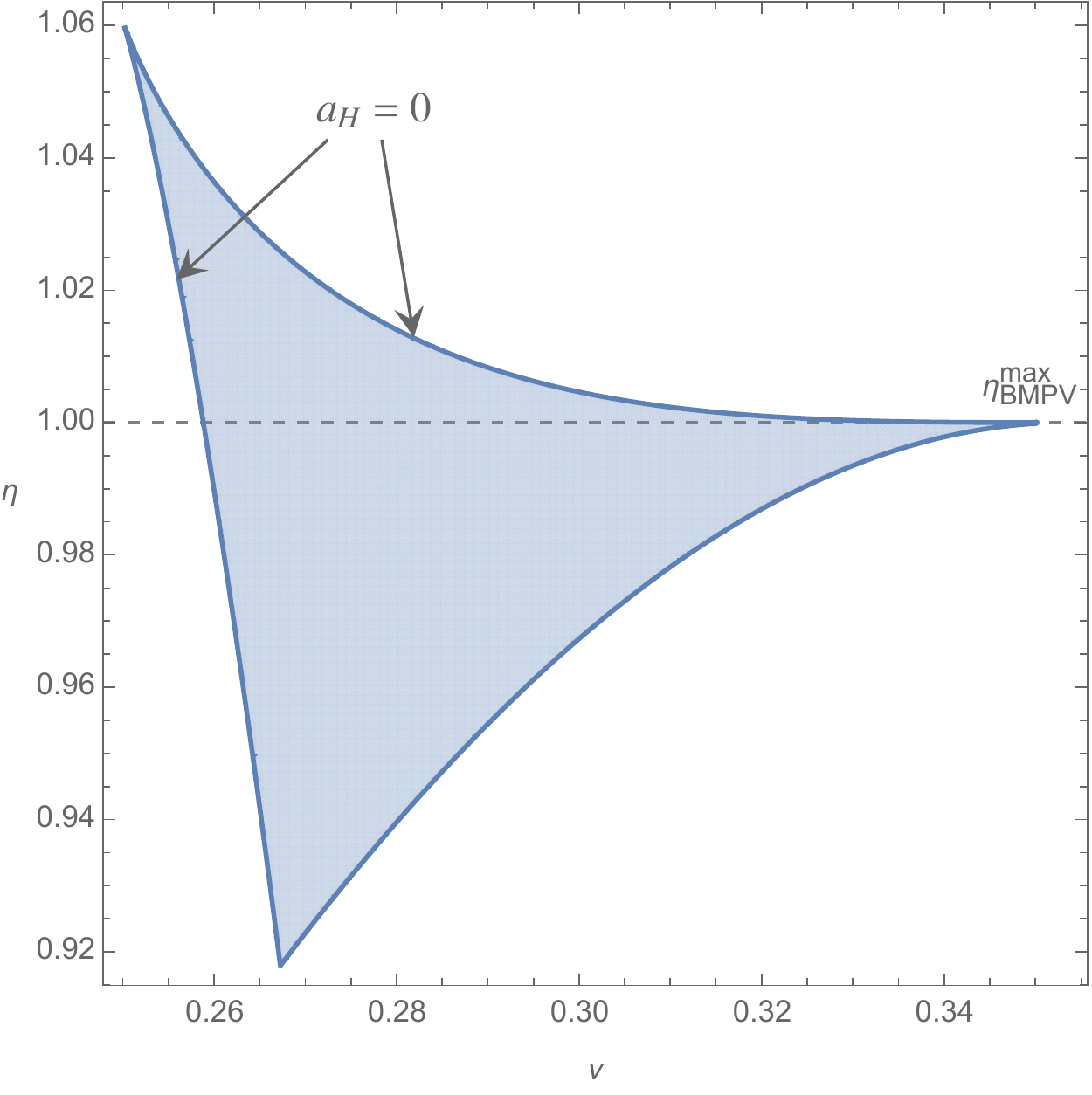}
\caption{}
\label{zeroJphi_III_region}
\end{subfigure}\hfill
\begin{subfigure}[t]{.48\textwidth}
\includegraphics[width=\textwidth]{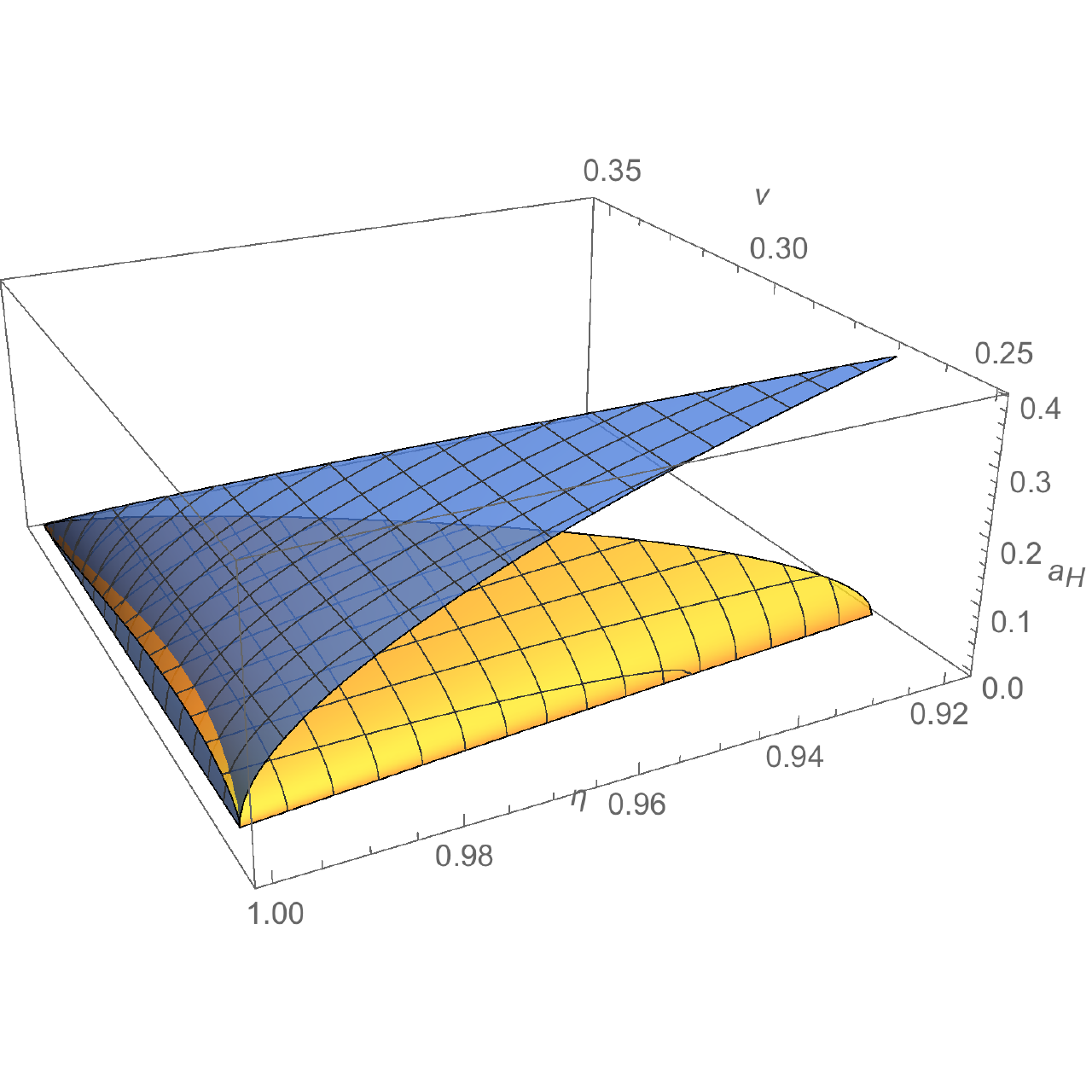}
\caption{}
\label{zeroJphi_III_3darea}
\end{subfigure}
\caption{\subref{zeroJphi_III_region} Moduli space for the $J_\phi=0$ spherical black hole III ($a_1<0<a_2$, $h_1=1$, $h_0=-1$,
  $h_2=1$). \subref{zeroJphi_III_3darea} Dimensionless area of the spherical black hole III (orange/lighter) and the BMPV black hole (blue/darker) within the region of overlap.  Again $a_H>a_{\text{BMPV}}$ in a narrow region close to $\eta=1$.}
\label{zeroJphi_III}
\end{figure}

We may now express the moduli space defined by \eqref{inequs_zeroJphi_III} in terms of the physical variables. 
We find this reduces to 
\begin{equation}
\begin{gathered}
 \frac{1}{4} < \nu < \frac{1}{2\sqrt{2}} ,\qquad \text{max} \left( \eta_-(\nu), \sqrt{2}(3 \nu- 8 \nu^3) \right)< \eta < \eta_+(\nu) \,,    \label{inequs_zeroJphi_IIIphys}
\end{gathered}
\end{equation}
where $\eta_\pm(\nu)$ are given again by \eqref{etapmSBH}, implying the range
\begin{equation}
  \begin{gathered}
   \frac{17}{7\sqrt{7}}<\eta < \frac{3}{2\sqrt{2}} .
  \end{gathered}
\end{equation}
The resulting moduli space is again a triangular region depicted in Figure \ref{zeroJphi_III_region}.
The reduced area is
\begin{equation}
  a_H = \left[
    \left(16 \nu ^2-1\right)^3
    - \left(\eta +6\sqrt{2} \nu  \left(8 \nu^2 - 1\right)\right)^2
  \right]^{1/2}\,.
\end{equation} 
Note that the upper/lower bounds $\eta_\pm(\nu)$ again arise from positivity of the area $a_H^2>0$.  The area of the black hole solution is plotted in Figure
\ref{zeroJphi_III_3darea}. It is clear that near $\eta=1$ this solution also can have higher entropy than the BMPV black hole.

This moduli space in this case is very reminiscent of that of
the spherical black hole I with $J_\phi=0$ described in section
\ref{sec:zeroJphi_I}. In fact, the expressions for the area $a_H$ as a function of $\eta$ and $\nu$ are identical for those two cases. 
However, the two moduli spaces \eqref{inequs_zeroJphi_I} and \eqref{inequs_zeroJphi_IIIphys} do not agree overall, as the remaining
inequalities, which determine the other part of the lower boundary curve, are not
equivalent for the two solutions. Furthermore it should be emphasised that the dipoles $\nu$ have a different meaning since the solutions have different spacetime topology (recall these are the magnetic potentials evaluated on a 2-cycle).   It thus appears to be a curious coincidence that the area functions for these two solutions are the same in this special case. Indeed, inspecting the area as a function of the physical charges for these two solutions with $J_\phi\neq 0$, \eqref{AreaSBHI}, \eqref{AreaSBHIII}, reveals that they are in fact distinct in general  (modulo the constraint \eqref{Jphi_I}).

Nevertheless, the similarity of the two solutions
strongly suggests that the spherical black hole III will have  a
soliton limit. Indeed, the top left point on the boundary of the moduli space in Figure \ref{zeroJphi_III_region}, which corresponds to the intersection of the upper and lower bounds arising from $a_H^2>0$, is again $(\eta,\nu)=(\frac{3}{2\sqrt{2}}, \frac{1}{4})$. This corresponds to the soliton solution \eqref{etanu_sol} in a different gauge, namely, the polar coordinates are adapted to the middle centre instead of the first centre. The solution to the constraints, together with
$J_\phi=0$, now gives
\begin{equation}
  k_2=k_1\,,\qquad
  a_2=-a_1=
\frac{k_1^2}{3}\,,
\end{equation}
and the charges are again given by \eqref{solitoncharges} (where now $C_1$ and $C_2$ are the 2-cycles corresponding to the axis rods $[a_1, 0]$ and $[0,a_2]$ respectively).

We now consider this family of black hole solutions near the soliton point. The calculation is identical to that for the spherical black hole I~\cite{Horowitz:2017fyg}. In fact, since the moduli space near this point is determined by the same boundaries curves, which correspond to the area vanishing, the expansion of the area of the black hole solution near the soliton point is the same as for the spherical black hole I, so we do not repeat it here.
Thus just as for the spherical black hole I~\cite{Horowitz:2017fyg}, we may interpret this as the area of a small nonrotating extremal black hole sitting in the soliton geometry.

\section{Comparison of entropies}\label{sec:comparison}

We have established in the previous section that there are three solutions which have the same conserved charges as the
BMPV black hole and whose entropy near the BMPV bound,
$\eta = 1$, may exceed that of the BMPV solution. Naturally, we want to compare
the entropies of the different solutions in this region.

\begin{figure}[t]
\begin{subfigure}[t]{.48\textwidth}
\includegraphics[width=\textwidth]{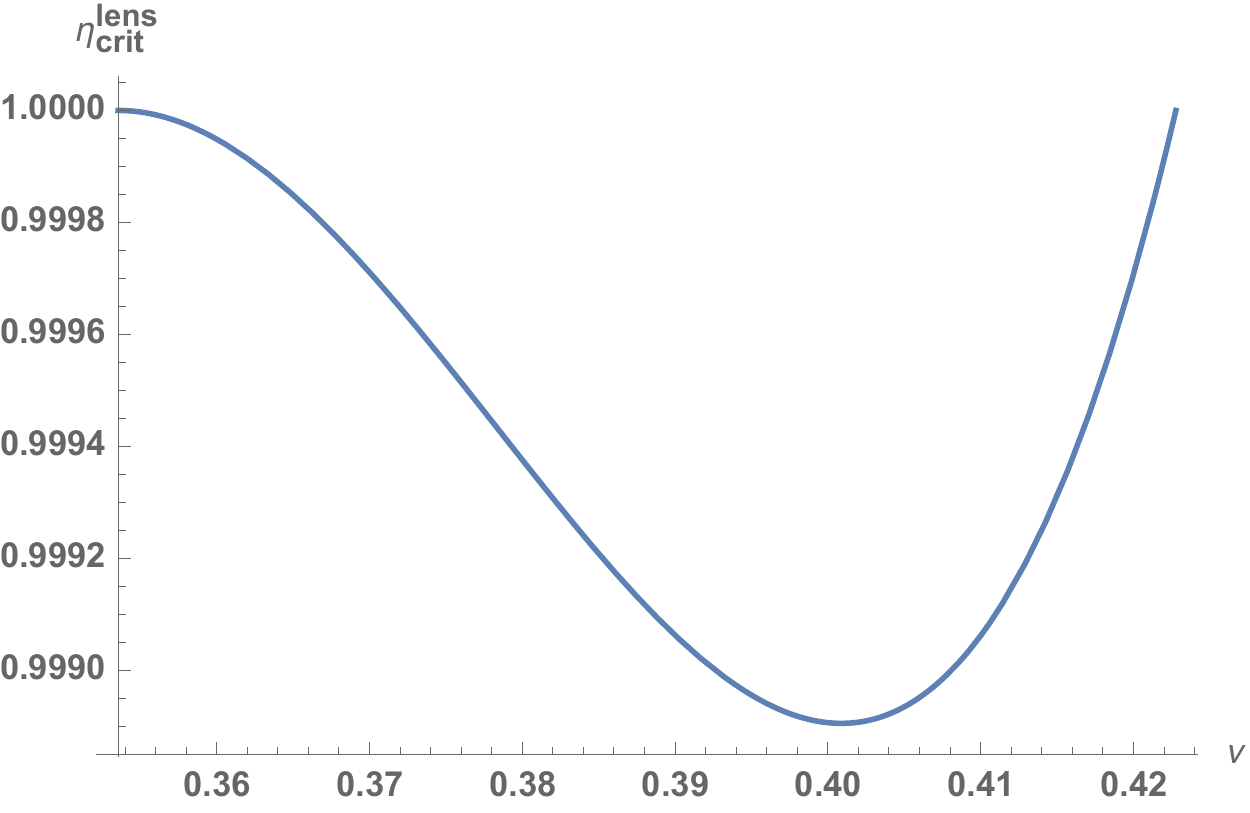}
\caption{}
\label{eta_crit_lens}
\end{subfigure}\hfill
\begin{subfigure}[t]{.48\textwidth}
\includegraphics[width=\textwidth]{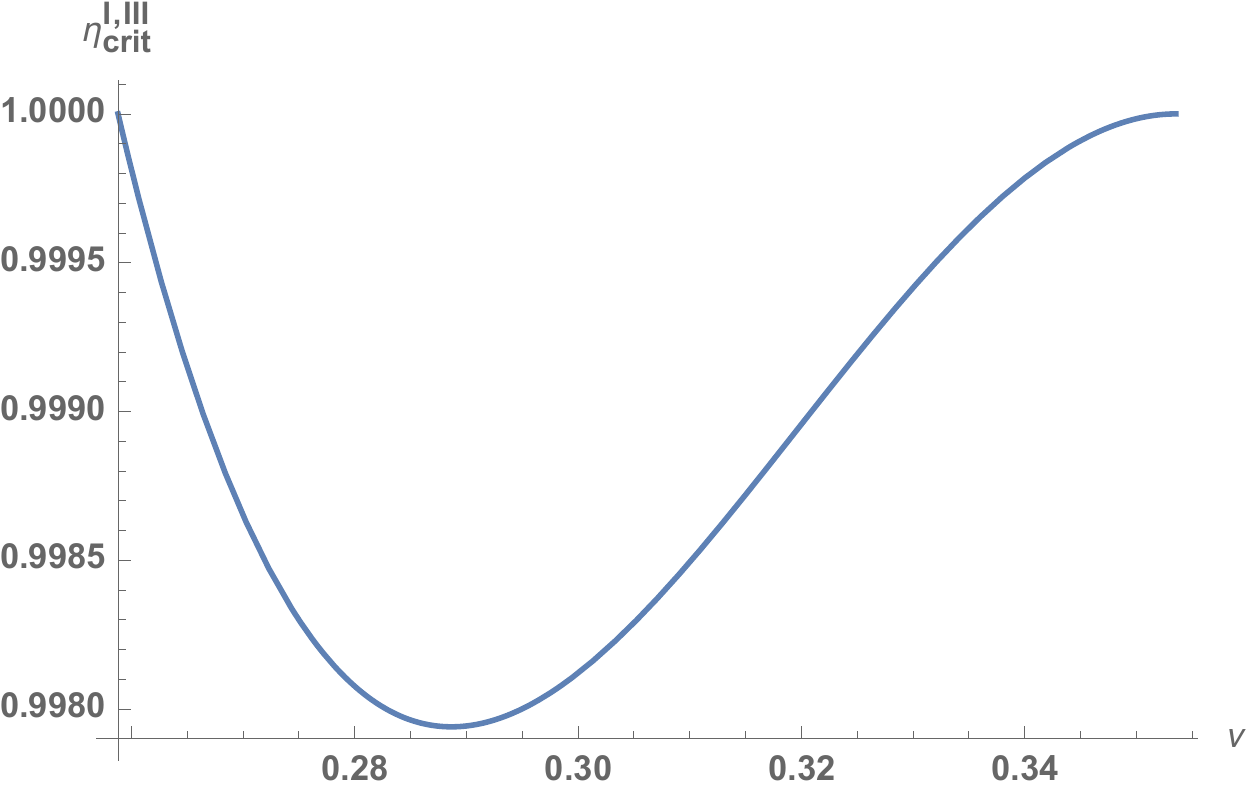}
\caption{}
\label{eta_crit_I_III}
\end{subfigure}
\caption{$\eta_{\text{crit}}$ for \subref{eta_crit_lens} the black
  lens and \subref{eta_crit_I_III} the spherical black holes I and
  III.}
\label{eta_crit}
\end{figure}

In particular, we are interested in determining the subregion of the moduli space with the same charges as BMPV, so $\eta<1$, for which the area exceeds that of the BMPV back hole, \ie $a_H \geq \sqrt{1-\eta^2}$.  We found that the spherical black hole I and III happen to have the same area function $a_H(\eta, \nu)$, so in both cases this subregion is given by
\begin{equation}
\eta \geq \eta_\text{crit}^{\text{I},\, \text{III}} \equiv \frac{1+ 20 \nu^2 - 32 \nu^4}{6 \sqrt{2} \nu} , \qquad  0.259\approx\frac{1}{2} \sqrt{2 - \sqrt{3}} < \nu <\frac{1}{2 \sqrt{2}}  \; .
\end{equation}
On the other hand, for the black lens we find it is given by
\footnote{The exact upper limit of $\nu$ is given by the unique positive real
  root of $-9+72\nu_0^2-144\nu_0^4+128\nu_0^6=0$.}
\begin{equation}
\eta \geq \eta_\text{crit}^\text{lens} \equiv \frac{1}{2\sqrt{2}} \left( \nu ( 9-8 \nu^2)  + (1-8\nu^2)\sqrt{2- 7 \nu^2} \right), \qquad \frac{1}{2 \sqrt{2}} < \nu <0.423  \; .
\end{equation}
It is worth noting that the curves $\eta_\text{crit}$ are both very
close to the BMPV upper bound $\eta=1$, see Figure \ref{eta_crit}. To emphasise, all spherical black hole I and III and black lens solutions with parameters in the above respective ranges possess an entropy greater than the BMPV solution.

We will now compute the maximum entropy solutions for fixed $\eta$. Since we are interested in the region where the new solutions dominate over the BMPV solution, which occurs very near $\eta=1$, it suffices to work in an expansion in $(1-\eta)$.  To find the maximum we need the appropriate root of $\partial_\nu a_H=0$ that gives a curve $\nu = \nu_*(\eta)$ along which $a_H$ is maximised for fixed $\eta$.  We find that near $\eta=1$ these are given by
\begin{align}
\nu^{\text{I},\, \text{III}}_*(\eta) &\approx 0.284+ 2.025( 1-\eta)\,, \qquad& a^{\text{I},\, \text{III}}_{H,\text{max}} &\approx 0.059+ 2.404(1-\eta)\,,  \\
 \nu^{\text{lens}}_*(\eta) &\approx 0.406 -  3.604 ( 1-\eta) , \qquad& a^{\text{lens}}_{H,\text{max}} &\approx 0.042+ 4.364 (1-\eta)  \,.
\end{align}
\begin{figure}[t]
\centering
\includegraphics[width=0.7\textwidth]{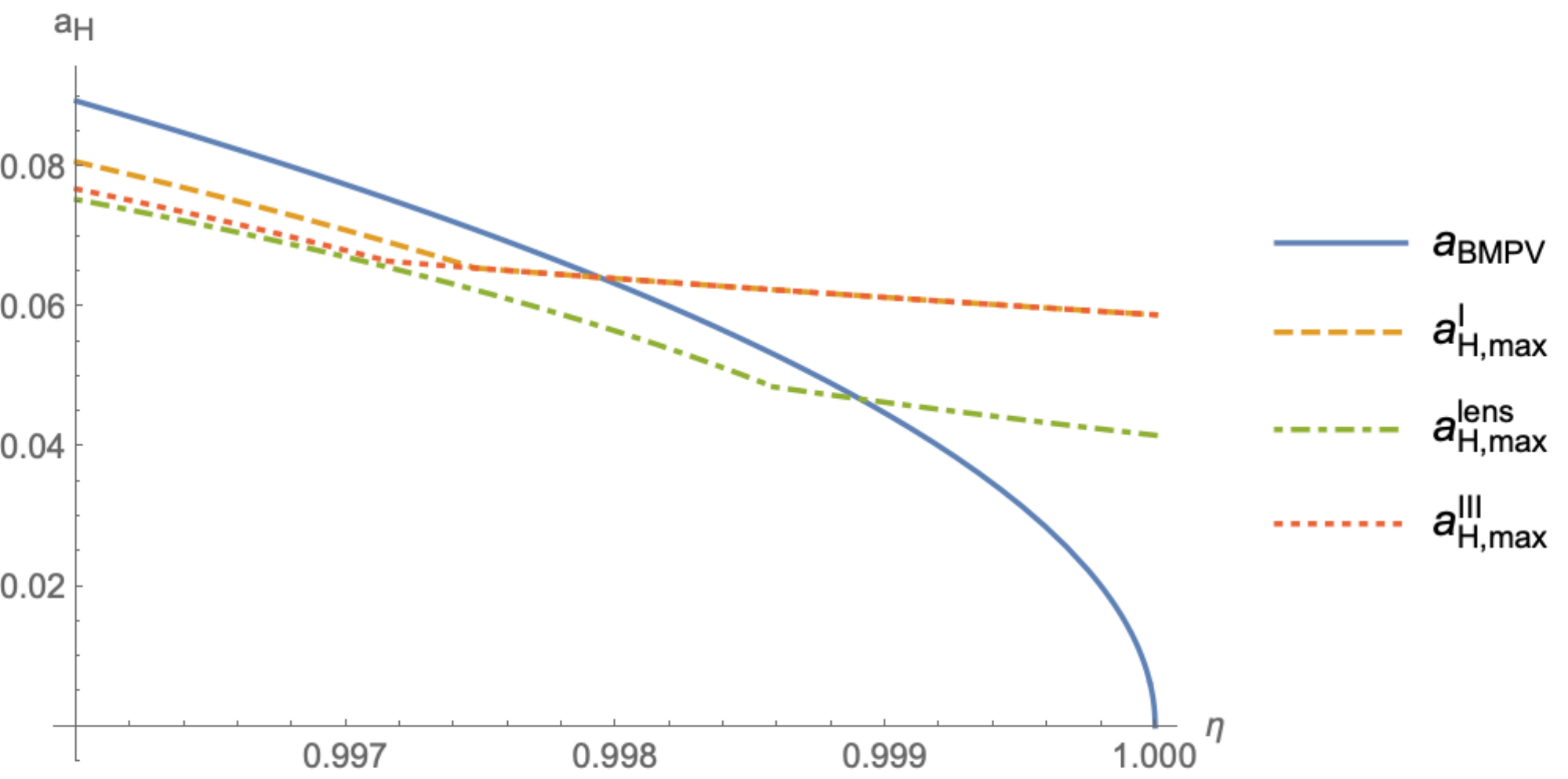}
\caption{Maximum horizon areas of the different solutions near
  $\eta=1$. Spherical black holes I and III have maximum areas exceeding those of the other solutions for $\eta >   0.997940$.}
\label{areaComparison}
\end{figure}%
The exact expressions for the maximum area are plotted in Figure
\ref{areaComparison}.  In particular, notice that near $\eta=1$, the
spherical black hole I and III have the same maximum area, which exceeds that
of BMPV for
\begin{equation}\label{aHI,III>aBMPV}
  0.997940 \approx \frac{11\sqrt{2}}{9\sqrt{3}} < \eta < 1\,.
\end{equation}
The black lens solution has a maximum area greater than that of BMPV for
\begin{equation}
0.998906 \approx \frac{37}{14\sqrt{7}} < \eta < 1\,,
\end{equation}
however its maximum area is always less than that of the spherical black hole I
and III solutions. Thus in the region \eqref{aHI,III>aBMPV} of the
moduli space the spherical black hole I and III are both equally
entropically favoured over the other solutions.

\section{Discussion}\label{sec:discussion}

In this work we have performed a detailed study of the moduli space of
asymptotically flat supersymmetric black holes in five-dimensional
minimal supergravity.  This was possible due to the recent
classification of such solutions under the assumption of biaxial
symmetry, which shows that these solutions must have a
Gibbons--Hawking base with harmonic functions of multi-centre
type~\cite{Breunholder:2017ubu}. For technical reasons we have only
fully analysed the three-centred single black hole solutions, although
this is rich enough to demonstrate several new features. In particular
we have focused on the solutions with the same conserved charges as
the original BMPV black hole and found new examples of black holes of
this type with both spherical topology and lens space $L(3,1)$
topology horizons. Furthermore, we have found that some of these new
solutions may possess greater entropy than the BMPV solution,
including a black lens, thus further adding to the ``single black hole
entropy enigma"~\cite{Horowitz:2017fyg}.

Curiously, we find that two of the spherical black holes with
nontrivial spacetime topology are equally dominant entropically in a
narrow region near the BMPV upper spin limit, one of which is the
previously known case~\cite{Horowitz:2017fyg}.  We emphasise though
that these may not be {\it the} maximum entropy black hole
states. There remains an infinite class of multi-centred solutions
with four or more centres which have not been studied, that could give
further examples of black holes with even greater entropy. Indeed,
from this perspective it is perhaps not even clear that a maximum
entropy state exists. This is because presumably there is no upper
bound on the number of independent 2-cycles in the exterior region
(which is determined by the number of centres) and it is expected that
adding 2-cycles increases the entropy as argued
in~\cite{Horowitz:2017fyg}.  It would be interesting to explore this
further.
 
The main technical barrier to analysing solutions with more than three
centres is to find an effective way of solving the constraints on the
parameters to obtain a useful description of the moduli spaces. In
particular, the smoothness and causality conditions
\eqref{smoothness_causality} can impose extra inequalities on the
parameters which are hard to extract. For the three-centred solutions
studied in this paper we verified numerically that positivity of the
mass, together with the basic inequalities arising from smoothness at
the three centres (one of which is a horizon), in fact appear to imply
the smoothness and causality conditions.  It would be interesting to
investigate if this is also the case for the solutions with more than
three centres and if such a conjecture could be proven. If so, this
would give much greater understanding and control of the moduli
spaces.
 
It is interesting to note that the dominant entropy solutions we find
may be both interpreted as black holes sitting in a bubbling soliton
spacetime. This is because they both admit a limit in which the black
hole shrinks to zero size leaving a regular soliton spacetime.  For
three-centred solutions, there is only one soliton spacetime with
equal angular momenta, so in fact we find there are two different ways
of adding a black hole to this soliton. This corresponds to ``adding''
a black hole at the two inequivalent smooth centres of the soliton
(the third centre is related to the first by a discrete symmetry).
Thus, extrapolating to $n$-centred solutions, we may expect up to $n$
inequivalent $S^3$-black holes sitting in a given soliton spacetime
(for symmetric rod structures this number would be reduced).
 
However, it is worth emphasising that we also found examples of
spherical black holes with the same charges as BMPV, but less entropy,
which do not admit a soliton limit. Instead these exist for
arbitrarily small angular momentum and reduce in a limit to the
Reissner-Nordstr\"om solution. Thus they may be interpreted as a
nonrotating black hole ``dressed" with nontrivial topology in the
exterior. It is of course possible that these solutions also admit a
soliton limit in the general moduli space where the angular momenta
are not equal, although we have not analysed this.  In any case, it is
clear that the black lens solutions never have a soliton limit.

There remains the puzzle of why the original counting of microstates
gave the entropy of the BMPV black hole~\cite{Breckenridge:1996is}.
The results of our work further complicate the picture found
in~\cite{Horowitz:2017fyg}. Not only are there other spherical black
holes dressed with topology, but also black lenses with greater
entropy than the BMPV black hole! Indeed, our results hint that near
the BMPV upper spin limit there could be many more solutions with
greater entropy than the BMPV black hole. Distinguishing and
identifying these BPS states in the string and brane system, or within
the CFT dual to their AdS$_3\times S^3$ decoupling limit, is clearly
an important problem.\\

\noindent {\bf Acknowledgements}. VB is supported by an EPSRC
studentship.  JL is funded in part by STFC [ST/L000458/1].

\appendix

\section{Smoothness and causality in the domain of outer communication}

As outlined in section \ref{sec:form_of_solutions}, for a solution to
be smooth and stably causal in the DOC, we require
\begin{equation}\label{SmoothnessAndCausality}
  K^2+HL > 0\,, \qquad g^{tt}<0\,.
\end{equation}
The smoothness condition $K^2+HL>0$ is (away from the centres) equivalent to
\begin{multline}
  \label{smoothness}
  \tilde{I} \equiv r^2 r_1 r_2(K^2+HL ) = r_1r_2h_0\ell_0 -
  r^2h_1h_2(h_1k_2-h_2k_1)^2 + rr_2h_1(\ell_0-h_0k_1^2) \\+
  rr_1h_2(\ell_0-h_0k_2^2)+ rr_1r_2h_0+r^2r_2h_1+r^2r_1h_2 > 0\,.
\end{multline}
The left hand side of this is in general a complicated function on the
2-dimensional orbit space $r>0$, $0\leq \theta \leq \pi$. 

In fact, for the black lenses with $h_0=3$, $h_1=h_2=-1$, we can show
that \eqref{smoothness} is automatically satisfied as a consequence of \eqref{conds}--\eqref{inequs_centres} as
follows. We can exploit the inequalities 
\eqref{inequs_centres} to obtain
\begin{multline}\label{Itilde}
  \tilde{I} \geq 3 r_1 r_2 
 \ell_0 + \left(
    -r+\frac{ r_2 |a_1|+ r_1 |a_2|}{|a_2-a_1|}\right) r(k_1-k_2)^2 + 
    r\big( r_2|a_1| + r_1 |a_2| + 3 r_1r_2 - r r_2 - r r_1\big)\,.
\end{multline}
Since $h_0 \ell_0 = 3\ell_0>0$ is implied by positivity of the area of
the horizon \eqref{inequ_horizon}, the first term in \eqref{Itilde} is always strictly positive.
Furthermore, from basic triangle inequalities,
\begin{equation}
  r_2|a_1| + r_1 |a_2| + 3 r_1r_2 - r r_2 - r r_1 
  = r_1 (r_2-r+|a_2|) + r_2 (r_1-r+|a_1|) + r_1 r_2 \geq r_1 r_2 > 0\,,
\end{equation}
showing that the third term is always positive.
Lastly, one can show that
\begin{align}
  -r|a_2-a_1|+ r_2 |a_1|+ r_1 |a_2| \geq 0\,,
\end{align}
so the second term is nonnegative. Therefore $\tilde{I}>0$ everywhere away from the centres, so the smoothness of the spacetime is guaranteed
without further restrictions. 

For the other 3-centred solutions it is not as straightforward to
prove the smoothness condition. In general establishing $K^2+HL>0$ requires input from
the full condition for positivity of the horizon area,
$h_0\ell_0^3 - h_0^2 m_0^2 >0$,  the left hand side of which will be a
complicated higher order polynomial in the remaining parameters once
the constraint equations \eqref{conds} are solved for $\ell_0$ and
$m_0$. Furthermore, in all cases we have been unable to prove the causality condition $g^{tt}<0$. Therefore, for the remaining cases we have performed numerical checks
that the smoothness and causality conditions \eqref{SmoothnessAndCausality} are satisfied as a consequence of \eqref{conds}--\eqref{inequs_centres} and \eqref{posmass}, as follows.

Since we know
$K^2+HL>0$ holds sufficiently far from the centres, we have checked
the condition for $10^4$ randomly chosen points within a region of
radius $r_\text{max} = 3\max(|a_1|,|a_2|)$ around the origin $r=0$
(the position of the horizon). We have done this for a
set of $10^4$ randomly chosen parameters satisfying
\eqref{conds}--\eqref{inequs_centres} and \eqref{posmass} for
each of the seven general solutions listed in Figure \ref{rod_structures} {\it and} the four special cases with equal angular momenta listed in Figure \ref{table}. The numerical checks confirm that in all cases the smoothness condition is satisfied without further
restrictions on the parameters. In a similar manner ($10^3$ points for
$10^4$ parameters each) we have also checked
that the causality condition $g^{tt}<0$ holds without further restrictions on the
moduli space. 

This provides evidence for the following conjecture: \eqref{conds}--\eqref{inequs_centres}
together with \eqref{posmass} imply \eqref{SmoothnessAndCausality} is automatically
satisfied.

%\printbibliography
\bibliographystyle{jhep}
\bibliography{bibl}

\end{document}